\documentclass[a4paper,11pt]{article}
\usepackage{jheppub} 

\usepackage{lineno}

\usepackage{tikz}
\usetikzlibrary{decorations.markings}
\usepackage{mathrsfs}
\usepackage{amssymb}

\usepackage{graphicx}
    
\usepackage{verbatim}

\usepackage{subcaption}

\usepackage{marginnote}

\usepackage{braket}
\usepackage{slashed}

\usepackage{textgreek}
\newcommand\varpsi{\scalebox{1}[.85]{\textsl{\textpsi}}}

\makeatletter
\newsavebox{\@brx}
\newcommand{\llangle}[1][]{\savebox{\@brx}{\(\m@th{#1\langle}\)}%
  \mathopen{\copy\@brx\kern-0.5\wd\@brx\usebox{\@brx}}}
\newcommand{\rrangle}[1][]{\savebox{\@brx}{\(\m@th{#1\rangle}\)}%
  \mathclose{\copy\@brx\kern-0.5\wd\@brx\usebox{\@brx}}}
\renewcommand{\@fpheader}{\relax}
\makeatother

\usepackage{soul}

\title{\boldmath Spectral Spacetime Entropy for Quasifree Theories}

\author[a,b]{Joshua Y. L. Jones}
\author[a]{and Yasaman K. Yazdi}
\affiliation[a]{School of Theoretical Physics, Dublin Institute for Advanced Studies,\\
10 Burlington Road, Dublin 4, Ireland}
\affiliation[b]{School of Mathematics, Trinity College,\\
Dublin 2, Ireland}

\emailAdd{jjones@stp.dias.ie}\emailAdd{ykyazdi@stp.dias.ie}

\abstract{Motivated by the necessity to UV-regularise entanglement entropy, we present a spectral method for calculating the entropy of quasifree states, for both bosonic and fermionic field theories. This construction is defined in spacetime rather than on a hypersurface, enabling the covariant regularisation of entropies, and its calculation in generic spacetime regions. We derive these formulae, which have previously appeared in the literature, in a new manner and highlight certain aspects of them, such as their connection to the density matrix and its eigenvalues. The spacetime nature of the formulation makes it particularly apt in the context of semiclassical and quantum gravity and in connection to black hole entropy. Another useful property of the formulation is its application to settings where no notion of a Cauchy surface exists, such as in the causal set theory approach to quantum gravity. We show example applications of the formulae which demonstrate their ability to reproduce known results. We also show a calculation in a causal set in $1+1$ dimensions which makes use of several of the unique and useful features of the formalism. In this last example, we obtain a novel result of a slightly modified entanglement entropy scaling coefficient, giving a possible signature of spacetime discreteness.}

\begin{document}
\maketitle
\flushbottom

\section{Introduction}

While it is not known which mathematical structures used in physics today will continue to exist in full quantum gravity,
we can still reason about the mathematical structures valid in the preceding intermediate regimes. One such regime is that of quantum field theory in curved spacetime, in which quantum matter fields are considered on a fixed smooth background spacetime described by general relativity.

As we approach higher energies, it is likely that we will reach a point where a smooth classical background spacetime is no longer adequate to describe the effects of gravity on matter, while the general Hilbert space formalism of matter fields remains valid.\footnote{Many issues with quantum fields occur at small distances where quantum field theory in the continuum breaks down, due to unphysical divergences stemming from UV-incompleteness. A discrete spacetime resolves these issues without needing to do away with the conventional formalism of quantum field theory.} At this stage, it is reasonable to replace the smooth background spacetime with something more fundamental. We would like this more fundamental structure to still exhibit general covariance or frame independence, an essential feature of general relativity with stringent bounds on its violation \cite{GRTest1, GRTest2}.  It thus seems that at least in this intermediate regime, a conservative, yet principled approach to describing the dynamics of matter fields in quantum gravity is a covariant operator formalism expressed in terms of a Hilbert space that can be defined on a more general spacetime structure.

Another important guide in bringing together quantum field theory and general relativity is the ubiquitous notion of information \cite{Shannon1948}. All physical theories have degrees of freedom, and understanding those degrees of freedom both qualitatively and quantitatively is key to gaining more insight into these theories. Similarly, the information carried by the degrees of freedom plays an important role in understanding physical processes. Several tools from information theory have turned out to be useful in the context of quantum field theory in curved spacetime and beyond. Entropy, which is perhaps the most intuitive and natural measure of information, has played an especially central role in this regard.

Indeed, one of the major open questions that we hope a theory of quantum gravity will answer is: what is the microscopic nature and origin of the Bekenstein-Hawking entropy of black holes \cite{Bekenstein72, Hawking75}? This question arose directly from informational considerations,  and its resolution is also likely to involve information in an essential way. A similar entropy exists in the presence of other (e.g. cosmological and Rindler) causal horizons \cite{MassarParentani2000, JacobsonParentani2003}, and is proportional to the event horizon area in the same way. This universality gives us an important clue regarding the nature of the degrees of freedom and information responsible for the entropy. Namely, the degrees of freedom should be of the same general type in all of these cases. Together with the observations mentioned in the earlier paragraphs, this aim for generality is one of the main motivations for the work we present and develop in this paper.

Specifically, we will consider a spectral formulation of entanglement entropy, suitable for applications to quasifree quantum field theories\footnote{We will use the phrase quasifree theory to denote a field theory together with a  quasifree state.} in \emph{spacetime} (or a suitable replacement thereof), in contrast to the  common spatial treatment; due to its spacetime nature, it is crucially also amenable to covariant regularisation. We will return to this point shortly.  In conventional calculations of entanglement entropy, a hypersurface is selected, upon which a state is specified via a density operator, $\rho$. One then obtains a reduced operator, $\rho_{red}$, by  tracing out inaccessible degrees of freedom, and calculates the entropy associated to it using the von Neumann formalism, $S = -\text{Tr}\left( \rho_{red} \ln\rho_{red}\right)$ \cite{Sorkin:1984kjy}. This entanglement entropy is well known to diverge when calculated on a smooth Lorentzian manifold without regularisation. After imposing an ultraviolet (and sometimes also an infrared) regularisation, an area law scaling is obtained. The area in the (spatial) area scaling of the entanglement entropy is the area of the boundary (in units of the regularisation) between two causally disconnected regions that are entangled. In particular, in spacetimes with an event horizon, this area would be the area of the event horizon. This close relation between entropy and area, shared by both horizon entropy and entanglement entropy, is one of the reasons why entanglement entropy has emerged as a leading candidate for the source of horizon entropy \cite{BKLS86,Solodukhin2011,Bianchi2013BHS,STwoHdof}.

As mentioned above, entanglement entropy is often defined and calculated on a spatial hypersurface. It is hence also typically regularised via a minimum length (or area) scale on this hypersurface. This spatial regularisation inevitably enters the final value of entropy, and due to being defined with respect to a particular frame,\footnote{Any minimum length or (codimension $\geq 1$) area scale is associated with a frame.} cannot ensure a covariant result. It is then clear that a more physical regularisation of entanglement entropy, if one wishes to relate it to the frame independent Bekenstein-Hawking entropy, must be a covariant one, regulated in a way that is unambiguous and frame independent. One way to ensure this is to regulate via a minimum spacetime volume. For this purpose, a notion of entropy defined in spacetime regions is necessary.

We will present such a spacetime formulation of entropy for bosonic and fermionic theories in Sections \ref{bosonsec} and \ref{fermsec} respectively, after some preliminaries in Section \ref{prel}. This entropy will be defined spectrally via an eigenvalue equation in terms of operators defined in  spacetime regions. When the operators are restricted to a subdomain (with non-empty causal complement) of the full spacetime region in which they are defined, the resulting entropy is an entanglement entropy. In Appendix \ref{examplesec}, we will show example calculations of thermal entropy in Minkowski spacetime and in Section \ref{sec: ee example} we will show a calculation of entanglement entropy in a discrete Rindler spacetime. 

This construction has a wide scope. As long as one can form a Hilbert space for fields, or calculate the two-point correlation function, one can apply the formalism. In particular, this entropy formulation can be used in any spacetime where the field theory has retarded and advanced Green functions (e.g. in globally hyperbolic spacetimes). It also applies to theories with a strictly global state definition (which is more general than the typical case where states can be localised to a hypersurface), and settings where the background has no differentiable metric, such as discrete spacetimes. This makes the construction particularly appealing in the context of quantum gravity, where one might consider replacing the smooth background with a more fundamental covariant structure, as mentioned earlier. In addition, this formulation has some practical, i.e. numerical, advantages in otherwise nontrivial settings. 
We further discuss these points in Section \ref{sec: ee example}. 

In Appendix \ref{ap: SJ} we discuss the Sorkin-Johnston state. This vacuum state definition has similar properties to the spacetime definition of entropy and it is a natural pure state that pairs with it. Finally, we point out that the construction we present was previously  derived for scalar fields by Sorkin \cite{Sorkin2012}. Our presentation will be different, focusing on the Hilbert space rather than the phase space, and we extend the treatment to fermions, which also brings out a nice relationship between the spin-statistics and entropy. We include a derivation more in line with Sorkin's, for both bosons and fermions, in Appendix \ref{algebrasec}.

\section{Preliminaries}\label{prel}

We begin by reviewing some conventions, definitions, and relevant background. We then review some key aspects of quasifree states, which we will use throughout the main text.\

\subsection{Notation and Conventions}

In general, we will work with a time-oriented smooth Lorentzian manifold $\mathcal{M}$, with metric $g$, and spin structure.\footnote{One may also take another background of choice (e.g. a nonsmooth structure such as a lattice), as we do in Section \ref{sec: ee example}, provided one can define bosonic and fermionic quantum fields on it.} We take this manifold to be globally hyperbolic, or else, assume the existence of Green functions. We will use the ``mostly plus" metric convention, $(-,+,\ldots,+)$, and use Planck units. We will later show some example calculations in Minkowski spacetime, where we will denote the standard Minkowski inner product as
\begin{equation}
    p \cdot x := \sum_{\mu}\sum_{\nu}\eta_{\mu \nu} p^{\mu} x^{\mu}, \qquad \eta_{\mu \nu} = \text{diag}(-1,+1,\ldots,+1).
\end{equation}
Spatial vectors will be  denoted with an arrow above a letter, and spacetime vectors without an arrow. Typically, we will not place hats on  operators. We will denote general classical bosonic and fermionic fields that satisfy their respective equations of motion with $\varphi$ and $\varpsi$, and their quantum counterparts with $\phi$ and $\psi$. When writing an equation that holds for both species of (quantum) fields, we will use $\Phi$.

The scalar fields that we will work with satisfy the Klein-Gordon equation, given by
\begin{equation}\label{eq: KGeqn}
    (-\Box + m^{2})\varphi=0, \qquad \Box:= \nabla_{\mu} \nabla^{\mu},
\end{equation}
where $\nabla$ is the covariant derivative. When working with fermions, we will take the gamma matrices to have the standard Clifford algebra relation, namely
\begin{equation}
    \{\gamma^{\mu}, \gamma^{\nu}\} = 2 g^{\mu \nu},
\end{equation}
and we choose the Dirac adjoint to be an involution given by
\begin{equation}
    \bar\psi = -i\psi^{\dagger}\tilde{\gamma}^{0},
\end{equation}
where $\tilde{\gamma}^{0}$ is the frame gamma matrix, that is, it has the relation $(\tilde{\gamma}^{0})^{2}=-1$, and is related to $\gamma^{\mu}$ via the usual vielbein formalism. Our choice of gamma matrix relations means that the Dirac equation, the equation of motion for fermionic fields, is
\begin{equation}\label{eq: Diraceqn}
    (-\slashed{\nabla}+m)\varpsi=0,
\end{equation}
where we use the slash to denote contraction with the gamma matrices, that is $\slashed{v}:= \gamma^{\mu} v_{\mu}$. We will suppress fermion indices throughout the paper, hence summation over this index should be inferred in inner products, and when operators act on fermion fields.

In what follows, we will usually work directly with the Hilbert space formulation, centred around the density operator, although in Appendix \ref{algebrasec} we provide a treatment that is centred around the correlation functions of the field, more in the style of Wightman and general algebraic quantum field theory. These two treatments are equivalent. 

We will consider quasifree (or Gaussian) states, which  have a particular type of Hilbert space that we describe in the next subsection.

\subsection{Quasifree States}

We will work with free bosonic and fermionic theories, that is, theories with canonical commutation and anticommutation relations respectively \cite{Derezinski2006}. In their spacetime form, these are \cite{Peierls:1952cb} 
\begin{align}\label{eq: iDelta commutator}
  i \Delta(x,x'):=  [\phi(x),\phi(x')] = i(G_{R}(x,x') - G_{A}(x,x')),
\end{align}
and
\begin{align}\label{eq: iDeltaF anti-commutator}
 i \Delta_F(x,x'):=    \{\psi(x),\bar{\psi}(x')\} = i(G_{R}(x,x') - G_{A}(x,x'))
\end{align}
respectively, where $x$ and $x'$ are spacetime points (not necessarily of equal time). $G_{R}$ and $G_{A}$ are the retarded and advanced Green functions, respectively. For an equation of motion specified by a differential operator $K$, with the physical examples we are interested in being the Klein-Gordon (\ref{eq: KGeqn}) and Dirac equations (\ref{eq: Diraceqn}), these Green functions satisfy 
\begin{equation}\label{eq: gr ga}
    K_{x} G_{R,A}(x,x') = \frac{\delta^d(x-x')}{\sqrt{-g(x')}},
\end{equation}
where $d$ is the spacetime dimension. We can therefore see from \eqref{eq: iDelta commutator} and \eqref{eq: iDeltaF anti-commutator} that $\Delta$ and $\Delta_F$ satisfy the equation of motion. If $K$ is Green-hyperbolic, as the Klein-Gordon and Dirac operators are, $G_{R,A}(x,x')$ always exist and are unique in globally hyperbolic spacetimes. They can also be specified on manifolds that are not globally hyperbolic via the addition of boundary conditions (see e.g. \cite{AID78, Wald80}). The retarded Green function has support only if $x'$ is in the causal past of $x$, while the advanced one has support only if $x'$ is in the causal future of $x$. 

In conventional quantum field theory, states are vectors in a Hilbert space.  We will consider Hilbert spaces that are Fock spaces.  In the case of free theories, one can form such a Hilbert space by starting from the space of solutions of the classical equation of motion, and by a choice of subspace or positive definite inner product, arrive at an appropriate one-particle Hilbert space. This can also be expressed as a choice of complex structure over the classical solution space. The Fock space is then the direct sum of tensor products of the one-particle Hilbert space, i.e., the direct sum of the different particle number Hilbert spaces.  The field operator on this Fock space is the linear combination of the basis forming the one-particle Hilbert space and its complex conjugate, with the associated creation and annihilation operators as coefficients.

The field operator has an expansion in terms of an orthonormal basis of the one-particle Hilbert space. It is formed of classical solutions $\{f_j\}$, furnished with creation and annihilation operators to create a symmetric, or antisymmetric, Fock space, in the respective cases of bosons and fermions. The field operator can thus be written
\begin{equation}\label{eq: field expansion}
    \Phi(x) = \sum_{j} \left( a^{\dagger}_{j} f^{*}_{j}(x) + a_{j} f_{j}(x) \right).
\end{equation}
Here we have written the expansion for a single real scalar or Majorana fermion field \cite{ElliottFranz}. As we will primarily work in compact spacetimes, the basis of solutions, $\{f_j\}$, will be countable and a summation over a discrete index (as above) is appropriate. In an infinite volume, the basis can be uncountable, and one must in such a case use an integral together with the corresponding integral measure. We will replace the sums with integrals, where necessary. For the real scalar field, $f_{j}\in C_{\text{sc}}^{\infty}(\mathcal{M})$, while for the Majorana fermion, $f_{j}\in C_{\text{sc}}^{\infty}(D\mathcal{M})$, where $D\mathcal{M}$ is the spinor bundle \cite{Dimock1982DiracQF}. Here, the subscript $\text{sc}$ denotes spatial compactness (necessary for the hypersurface inner products to be well defined), that is, compact intersection with any Cauchy surface. The functions $f_{j}$ are solutions to the equation of motion, $Kf_{j}=0$, and as such can be written as the application of $\Delta$ or $\Delta_{F}$ to a test function with compact spacetime support. In the scalar case $\Delta$ acts on test functions in the space $C^{\infty}_{0}(\mathcal{M})$, and in the fermion case $\Delta_{F}$ acts on test spinor fields in $C^{\infty}_{0}(D\mathcal{M})$, to obtain solutions to $K$. In general, we will use $f^{*}_{j}$ to denote the charge conjugate solution to $f_{j}$. General bosonic or fermionic fields can be decomposed in terms of these fields, as one is not concerned about the field's transformation properties for the purposes of entropy calculation.\footnote{Although one should keep in mind the new form of any constraints on the decomposed fields.}

The algebraic commutation and anticommutation relations of the creation and annihilation operators are
\begin{align}
    [a_i,a_j^{\dagger}] = \delta_{ij}, \\
    \{a_i,a_j^{\dagger}\} = \delta_{ij},
\end{align}
for bosons and fermions respectively. We will consider the case of quasifree states, denoting the state as $|\Psi\rangle$. These are states for which the odd $n$-point correlation functions vanish, and the even $n$-point correlation functions can be written as sums of products of two-point correlation functions via Wick's rule, in the same manner as in a free theory. We thus must have
\begin{equation}\label{eq:wick}
    \scalebox{0.92}{$\displaystyle \langle \Psi | \Phi(x_{1}) \cdots \Phi(x_{n})|\Psi\rangle = \sum_{j} (\pm1)^{P_j} \langle \Psi | \Phi(x_{j_1(1)}) \Phi(x_{j_1(2)})|\Psi\rangle \cdots \langle \Psi | \Phi(x_{j_\frac{n}{2}(1)}) \Phi(x_{j_\frac{n}{2}(2)})|\Psi\rangle,$}
\end{equation}
 where $n$ is even. The sum is over every way of partitioning the set of integers $\{1,\ldots,n\}$ into an unordered set of $\frac{n}{2}$ ordered pairs, where $j_{i}$ is the $i^{\text{th}}$ pair of the $j^{\text{th}}$ partition, having $j_{i}(1)<j_{i}(2)$ (i.e., each pair preserves the index ordering on the left hand side). In the above expression, $P_j$ refers to the parity of the permutation to go from the sequence $1,\ldots,n$ to $j_{1}(1),\ldots,j_{\frac{n}{2}}(2)$. For bosons, we have $(+1)^{P_j}$, while for fermions, we have $(-1)^{P_j}$.\footnote{For example, the four-point functions of a real scalar field $\phi$, and Majorana fermion field $\psi$, are
\begin{equation*}
\begin{split}
    \langle \Psi | \phi(x_{1}) \phi(x_{2}) \phi(x_{3}) \phi(x_{4})|\Psi\rangle \phantom{+} = \phantom{+} &\langle \Psi | \phi(x_{1}) \phi(x_{2})|\Psi\rangle\langle \Psi | \phi(x_{3}) \phi(x_{4})|\Psi\rangle \\
    + &\langle \Psi | \phi(x_{1}) \phi(x_{3})|\Psi\rangle\langle \Psi | \phi(x_{2}) \phi(x_{4})|\Psi\rangle \\
    + &\langle \Psi | \phi(x_{1}) \phi(x_{4})|\Psi\rangle\langle \Psi | \phi(x_{2}) \phi(x_{3})|\Psi\rangle,
\end{split}
\end{equation*}
and
\begin{equation*}
\begin{split}
    \langle \Psi | \psi(x_{1}) \psi(x_{2}) \psi(x_{3}) \psi(x_{4})|\Psi\rangle \phantom{+} = \phantom{+} &\langle \Psi | \psi(x_{1}) \psi(x_{2})|\Psi\rangle\langle \Psi | \psi(x_{3}) \psi(x_{4})|\Psi\rangle \\
    - &\langle \Psi | \psi(x_{1}) \psi(x_{3})|\Psi\rangle\langle \Psi | \psi(x_{2}) \psi(x_{4})|\Psi\rangle \\
    + &\langle \Psi | \psi(x_{1}) \psi(x_{4})|\Psi\rangle\langle \Psi | \psi(x_{2}) \psi(x_{3})|\Psi\rangle
\end{split}
\end{equation*}
respectively.}

For a mixed state, in order to have Wick's rule hold, we must have a density matrix $\rho$ that has an entanglement (or modular) Hamiltonian \cite{EntanglementHamiltonian} that is quadratic in creation and annihilation operators. That is, provided that one regularises, e.g. by implementing an ultraviolet cutoff,\footnote{i.e. one regularises to have a type I algebra, and normal state \cite{KadisonRingroseII}, the physical situation wherein the entropy is well defined \cite{BKLS86}.} the density matrix should be expressible as
\begin{equation}\label{eq: rhoboson}
    \rho = \frac{1}{Z}e^{- \sum_{i,j} H_{i j} v^{\dagger}_{i} v_{j}},
\end{equation}
where $H_{ij}=H_{ji}^{*}$, $Z$ is a normalisation such that $\text{Tr}(\rho)=1$, and $v$ is a vector of creation and annihilation operators associated to a basis of the one-particle Hilbert space, or equivalently, a vector of field operators. This density matrix then has the property that $\text{Tr}(\rho\, O) = \langle O \rangle$, for any $O$ constructed out of field operators.

We can perform a basis transformation to diagonalise this operator, via the diagonalisation of $H$. This is always possible, as the density matrix is self-adjoint on the Hilbert space. This diagonalisation must preserve the canonical commutation or anticommutation relations, and as such is a Bogoliubov transformation of our initial chosen Hilbert space basis.

What this means is that we have a Fock space, formed of a one-particle Hilbert space that has some basis $\{f_{j}\}$. The density matrix must be diagonal in this basis, and have the form 
\begin{equation}\label{eq: rho diag}
    \rho = \prod_{j}\frac{e^{- \tilde{H}_{j j} a^{\dagger}_{j} a_{j}}}{Z_{j}},
\end{equation}
where $Z_{j} = \sum\limits^{\infty}_{k_{j}=0} e^{- \tilde{H}_{j j} k_{j}}$ , $\tilde{H}$ is the diagonalised entanglement Hamiltonian, and $k_j$ is the occupancy of the  $j^{\text{th}}$ mode. The creation and annihilation operators in \eqref{eq: rho diag} can correspond to those in \eqref{eq: field expansion}, since we can choose the $\{f_j\}$ to diagonalise $\rho$.  Note that since $\rho$ is positive semidefinite, self-adjoint, and has the property $\text{Tr}(\rho)=1$, the $\tilde{H}_{jj}$'s  are greater than 0. In this basis, using the expansion for the field operator \eqref{eq: field expansion} and considering some $n$-point function,
\begin{equation}
    \langle \Phi(x_{1}) \cdots \Phi(x_{n})\rangle = \text{Tr}\left( \rho\, \Phi(x_{1}) \cdots \Phi(x_{n})\right),
\end{equation}
one can show that Wick's rule must hold, via the insertion of the field operators, and the algebraic relations of the creation and annihilation operators.

Inspection of the density matrix \eqref{eq: rho diag} shows that it has a thermal form, for each mode, with ``temperature" $T=\frac{1}{\tilde{H}_{jj}}$. Accordingly, the pure state corresponding to the vacuum state coincides with the limit of $\tilde{H}_{jj} \to \infty$, as this leads to a nonzero contribution from the vacuum state only. Alternatively, we can write this state directly by defining the density matrix as a projection onto the vacuum state.

\section{Bosonic Entropy} \label{bosonsec}

A nonzero entropy is the signature of a system in a mixed state. In general, any mixed state can be obtained by tracing out part of a Hilbert space from a pure state defined in it, although that Hilbert space and pure state are not unique. Here we derive our entropy formula for a real scalar field in the case where the density matrix is diagonalisable in the sense of \eqref{eq: rho diag}. More complicated bosonic fields can be decomposed into real scalar degrees of freedom, before our formula is applied.

As we have seen, imposing that our state is Gaussian, ensures that there exists a basis for our state formed by the action of creation operators, $a^{\dagger}_j$, on the vacuum, and a density matrix that is quadratic in these operators, of a ``modewise thermal" form. 

Due to the diagonal density matrix, \eqref{eq: rho diag}, we essentially now have a classical probability distribution over all of the excited states for each mode, and can calculate the total entropy as the sum of the individual entropies of each mode. That is, we have
\begin{equation}\label{eq: full entropy}
    S_{j} = -\sum_{k_{j}=0}^{\infty} p_{j,k_{j}} \ln{p_{j,k_{j}}}, \qquad p_{j,k_{j}} = \frac{e^{- \tilde{H}_{j j} k_{j}}}{Z_{j}}, \qquad Z_{j} = \sum_{k_j=0}^{\infty} e^{-\tilde{H}_{jj}k_j},
\end{equation}
where $k_j$ is the occupancy of the  $j^{\text{th}}$ mode, as before. The total entropy is $S=\sum\limits_j S_j$.

 We now wish to express this entropy, for a bosonic theory, using the spacetime two-point correlation (or Wightman) function,
\begin{equation} \label{eq: Wightman}
    W(x,x') := \langle \Psi|\phi(x) \phi(x')|\Psi\rangle = \text{Tr}(\rho\, \phi(x) \phi(x')),
\end{equation}
which as we saw in the previous section, specifies all the higher $n$-point functions in a quasifree theory, and therefore specifies the full theory. Inserting into \eqref{eq: Wightman} the expansion of the field, \eqref{eq: field expansion}, in terms of the orthonormal basis that diagonalises the density matrix, we have
\begin{equation}
\begin{split}\label{Wexpr}
    W(x,x') &= \sum_{j} \sum^{\infty}_{k_{j}=0} \langle k |_{j} \frac{e^{- \tilde{H}_{j j} a^{\dagger}_{j} a_{j}}}{Z_{j}} \left(a^{\dagger}_{j} f_{j}^{*}(x) + a_{j} f_{j}(x)\right)\left(a^{\dagger}_{j} f_{j}^{*}(x') + a_{j} f_{j}(x')\right) |k\rangle_{j} \\
    &= \sum_{j} \sum^{\infty}_{k_{j}=0} p_{j,k_{j}} \left((k_{j}+1)f_{j}(x)f^{*}_{j}(x') + (k_{j}) f_{j}^{*}(x)f_{j}(x')\right),
\end{split}
\end{equation}
where $|k\rangle_{j}$ is the state corresponding to the $k^{\text{th}}$ occupancy of the $j^{\text{th}}$ mode (that is, $(a_{j}^{\dagger})^{k}|0\rangle$), and $\tilde{H}$ is the diagonalised entanglement Hamiltonian as before.

Let us make some observations regarding \eqref{eq: full entropy} and \eqref{Wexpr}. Our probability distribution can be expressed as a geometric series, allowing the identifications
\begin{align}
    S_{j}=-\sum^{\infty}_{k_{j}=0} p_{j,k_{j}} \ln p_{j,k_{j}} &=  A_j \ln A_j - B_j\ln B_j,\label{Deltaexpr}\\
     A_j&:=\sum^{\infty}_{k_{j}=0} (k_{j}+1) p_{j,k_{j}},\label{eq: A} \\  B_j&:= \sum^{\infty}_{k_{j}=0} (k_{j}) p_{j,k_{j}} = A_j-1,\label{eq: B} 
\end{align}
satisfying $A_j \geq 1$ and $B_j \geq 0$. We can show the above identifications more explicitly. We begin by writing the probability distribution of a single $j$-mode as
\begin{equation}\label{eq: probability of a mode}
    p_{j,k_{j}} = (1-e^{-\tilde{H}_{jj}})e^{- \tilde{H}_{j j} k_{j}}.
\end{equation}
The entropy of this mode is
\begin{equation}\label{eq: mode entropy}
    -\sum^{\infty}_{k_{j}=0} p_{j,k_{j}} \ln p_{j,k_{j}} =  \tilde{H}_{jj}\frac{e^{-\tilde{H}_{jj}}}{1-e^{-\tilde{H}_{jj}}} - \ln\left( 1-e^{-\tilde{H}_{jj}}\right).
\end{equation}
Inserting \eqref{eq: probability of a mode} into \eqref{eq: A} and \eqref{eq: B}, we obtain 
\begin{equation}\label{eq: ABj}
    A_j = \frac{1}{1-e^{-\tilde{H}_{jj}}}, \qquad B_j = \frac{e^{-\tilde{H}_{jj}}}{1-e^{-\tilde{H}_{jj}}},
\end{equation}
and so the combination on the right hand side of  (\ref{Deltaexpr}) is 
\begin{equation}
    A_j \ln A_j - B_j\ln B_j = \tilde{H}_{jj}\frac{e^{-\tilde{H}_{jj}}}{1-e^{-\tilde{H}_{jj}}} - \ln\left( 1-e^{-\tilde{H}_{jj}}\right),
\end{equation}
in agreement with \eqref{eq: mode entropy}. Note that the above calculations of the geometric and arithmetico-geometric series were possible due to the fact that $\tilde{H}_{jj}>0$.

Hence, we see that the essential quantities needed to calculate the entropy of each mode are the $A_j$ and $B_j$ coefficients of the two parts ($f_{j}(x)f^{*}_{j}(x')$ and $f_{j}^{*}(x)f_{j}(x')$) of the two-point function \eqref{Wexpr}. We subsequently sum the entropy for each mode to obtain the full entropy of the quantum state. As we have shown above, if we find the appropriate basis in which the density matrix is diagonal, we can calculate (via a Bogoliubov transformation) the two-point function modewise in this basis, and in this way extract  the $A_j$ and $B_j$ coefficients needed to calculate the entropy.

This gives a clear interpretation of the entropy of the field in a quasifree state. The entropy can be understood as the information in that state, given the joint probability distribution over the number of excitations of independent field modes. This is the natural way for the field to store information in a quasifree state.

However, the dependence of the above prescription on the basis that diagonalises the density matrix is inconvenient, as obtaining this basis can be very tedious. We therefore wish to cast the problem of finding the $A_j$ and $B_j$ coefficients in a basis independent form. This can be done by noticing that the commutator $i\Delta$ \eqref{eq: iDelta commutator}, expressed in terms of the field expansion \eqref{eq: field expansion}, is
\begin{equation}\label{eq: commutator in terms of field}
    [\phi(x),\phi(x')] = \sum_{j} \left(f_{j}(x)f^{*}_{j}(x') - f_{j}^{*}(x)f_{j}(x')\right).
\end{equation}
Comparing \eqref{eq: commutator in terms of field} to \eqref{Wexpr}, we see that the commutator has the same two parts as the (quasifree) two-point function, but with coefficients $1$ and $-1$. We can use this to solve the problem of finding $A_j$ and $B_j$ in a basis independent spectral manner in the following way:

\noindent We first promote the two-point function $W$ and commutator $i \Delta$ to operators via the duality (``adjoint") map $\mathscr{A}_{\llangle[\scriptscriptstyle]\cdot,\cdot\rrangle[\scriptscriptstyle]}: f \to \llangle f^{*}, \cdot \rrangle$, induced by $\llangle \cdot, \cdot \rrangle$, the inner product on the solution space with respect to which the $f_j$'s and $f^*_j$'s are orthonormal. Conventionally, the inner product used is the indefinite Klein-Gordon inner product, given by
\begin{equation}
    \llangle \varphi, \tilde\varphi\rrangle_{KG} := i\int_{\Sigma}\{ \varphi^*(x) n^{\mu}\nabla_{\mu} \tilde\varphi(x) - \tilde{\varphi}(x) n^{\mu}\nabla_{\mu} \varphi^{*}(x)\}\, dS,
\end{equation}
where $\Sigma$ is a Cauchy surface, $n^{\mu}$ is its unit normal, and $dS$ is the induced volume element on $\Sigma$. We then use this inner product to set up the generalised eigenvalue equation
\begin{equation}\label{eq: gen eig}
    \scalebox{0.9}{$\displaystyle \sum_{j} \left(A_{j} f_{j}(x)\llangle f_{j}, g \rrangle_{KG} + B_{j} f_{j}^{*}(x) \llangle f^{*}_{j}, g \rrangle_{KG} \right) = \lambda \sum_{j} \left(f_{j}(x) \llangle f_{j}, g \rrangle_{KG} - f_{j}^{*}(x)\llangle f^{*}_{j}, g \rrangle_{KG} \right).$}
\end{equation}
The generalised eigenfunction $g(x)$ and eigenvalue $\lambda$ solutions to \eqref{eq: gen eig} come in pairs, given by
\begin{align}
    g^{A}_{j}(x) = f_{j}(x), &\qquad \lambda^{A}_{j} = A_{j}, \label{eq: g and lambda A}\\
    g^{B}_{j}(x) = f^{*}_{j}(x), &\qquad \lambda^{B}_{j} = - B_{j} = 1-A_{j} = 1-\lambda^{A}_{j}   . \label{eq: g and lambda B}
\end{align}
Note, in particular, that each pair of eigenvalues $\lambda^{B}_{j}$ and $\lambda^{A}_{j}$ is equal to the $A_j$ and $B_j$ (the latter times $(-1)$) pair that we set out to obtain in a different manner.

Now we make an observation. The Klein-Gordon inner product can be recast in a spacetime form as \cite{Dimock1980}
\begin{equation}\label{eq: KG relation to L2}
\begin{aligned}
    \llangle \varphi, \tilde{\varphi}\rrangle_{KG} &= i\int_{\Sigma}\{ \varphi^{*}(x) n^{\mu}\nabla_{\mu} \tilde{\varphi}(x) - \tilde{\varphi}(x) n^{\mu}\nabla_{\mu} \varphi^{*}(x)\}\,dS \\
    &= -i\int_{\mathcal{M}}\varphi^{*}(x) h(x)\, dV =:-i\llangle \varphi, h \rrangle_{L^{2}}.
\end{aligned}
\end{equation}
Here both $\varphi$ and $\tilde\varphi$ are solutions to the equation of motion, and $\tilde{\varphi}:=\hat{\Delta} h$ (see \eqref{eq: Delta as operator}), with $h$ being a test function in $C^{\infty}_{0}(\mathcal M)$. Note that as $h$ is a test function, and not a solution, the right hand side is a pairing between two different spaces.

The objects that we have inside the Klein-Gordon inner product in \eqref{eq: gen eig}, in both $W$ and $i\Delta$, are solutions to the equation of motion. This means we can use \eqref{eq: KG relation to L2} to rewrite \eqref{eq: gen eig} to have the operators act on test functions defined over spacetime. To ensure that we have a one-to-one correspondence between the test functions satisfying the generalised eigenvalue equation and the solutions to the Klein-Gordon equation in \eqref{eq: gen eig} (and therefore the same number of generalised eigenvalues $\lambda$), we must ``mod out'' from our space components of $h$ that are in the kernel of $\hat\Delta$, leaving us with
\begin{equation}\label{eq: gen eig 2}
    \scalebox{0.94}{$\displaystyle \sum_{j} \left(A_{j} f_{j}(x)\llangle f_{j}, h \rrangle_{L^{2}} + B_{j} f_{j}^{*}(x) \llangle f^{*}_{j}, h \rrangle_{L^{2}} \right) = \lambda \sum_{j} \left(f_{j}(x) \llangle f_{j}, h \rrangle_{L^{2}} - f_{j}^{*}(x)\llangle f^{*}_{j}, h \rrangle_{L^{2}} \right),$}
\end{equation}
where $h \in C^{\infty}_0 (\mathcal M)/ \ker(\hat{\Delta})$. The generalised eigenvalues remain as before in this equation, and $h(x)$ is related to $g(x)$ via\footnote{We take $h$ to be some representative member of the equivalence class that maps to our previous $g$ under $\hat{\Delta}$.} 
\begin{equation}
    g(x) = \int_{\mathcal{M}}\Delta(x,x') h(x')\, dV'.
\end{equation}
More compactly, we can rewrite \eqref{eq: gen eig 2} as
\begin{equation} \label{Sev}
    \hat{W} h =i\lambda \hat{\Delta} h,
\end{equation}
where we have defined 
\begin{align}
    \hat{W}f &:= \llangle\Psi|\phi\, \mathscr{A}_{L^{2}} (\phi) |\Psi\rrangle f = \llangle\Psi|\phi\, \llangle \phi^{*}, f \rrangle_{L^{2}} |\Psi\rrangle, \label{eq: W as operator}\\
    i\hat{\Delta}f &:= [\phi, \,\mathscr{A}_{L^{2}}( \phi)] f = [\phi, \,\llangle \phi^{*}, f \rrangle_{L^{2}}],\label{eq: Delta as operator}
\end{align}
where $\mathscr{A}_{L^{2}}$ is the duality map induced by the $L^{2}$ inner product, $\mathscr{A}_{L^{2}}:f \to \llangle f^{*}, \cdot \rrangle_{L^{2}}$, and $f$ in \eqref{eq: W as operator} and \eqref{eq: Delta as operator} is an arbitrary function. We reiterate that in \eqref{Sev}, $h$ must not be in the kernel of $\hat{\Delta}$, i.e. $\hat{\Delta} h\neq 0$. Note that \eqref{Sev}, given all conventional choices, is
    \begin{equation}\label{eq: gen eig boson explicit}
        \int_{\mathcal{M}} W(x,x') h(x')\,dV' = i\lambda\int_{\mathcal{M}}\Delta(x,x') h(x')\,dV'.
    \end{equation}

Finally, we can obtain the full entropy of the field by summing over the $\lambda$'s in the generalised eigenvalue equation:
\begin{equation}
    S = \sum_\lambda \lambda \ln |\lambda|. \label{Ssum}
\end{equation}
Equation \eqref{Ssum} is equivalent to \eqref{eq: full entropy}, but by casting the problem as a generalised eigenvalue problem, \eqref{Sev}, we have provided ourselves with the freedom to solve the problem in any basis. In particular, we no longer need the basis that diagonalises $\rho$ in order to extract the $A_j$'s and $B_j$'s. We can set up the problem in any basis and solve it with the methods of generalised eigenvalue equations (including numerical methods). In Appendix \ref{appendix: algebraic S for bosons} we present a derivation of \eqref{Sev} and \eqref{Ssum} that proceeds in a more algebraic manner.

For pure vacuum states, all solutions to the generalised eigenvalue problem will have $\lambda=0$ or $\lambda=1$, resulting in a vanishing entropy from \eqref{Ssum}, as expected. We can see this by substituting $p_{j,0}=1$ and $p_{j,k\neq0}=0$ in \eqref{eq: A} and \eqref{eq: B}, or from the $\tilde{H}_{jj}\to\infty$ limit of \eqref{eq: ABj}, both of which would yield $A_{j}=1$ and $B_{j}=0$ (therefore $\lambda^A_j=1$ by \eqref{eq: g and lambda A} and $\lambda^B_j=0$ by \eqref{eq: g and lambda B}) and hence $S_{j}=0$. For maximally mixed states, $\tilde{H}_{jj}\rightarrow 0$, and the entropy diverges (c.f. \eqref{eq: mode entropy}). (We can also see that this corresponds to a mixed state where all excited states in our diagonal basis are equiprobable). If we start with a pure vacuum state, upon coarse-graining or a restriction to a subdomain with non-empty causal complement before solving \eqref{Sev}, the solution spectrum $\{\lambda\}$ will yield a nonzero (finite if UV regulated) entanglement entropy. Similarly, thermal states yield a nonzero entropy. We will show example calculations of these scenarios in Appendix \ref{examplesec}, and Section \ref{sec: ee example}.

The spectral formulation above has a number of other advantages. The operators $\hat{W}$ and $i\hat{\Delta}$ entering the generalised eigenvalue problem \eqref{Sev} have support in \emph{spacetime}. They hence lend themselves to spacetime regularisation, which is explicitly covariant, in contrast to spatial regularisation, which is not. Some type of regularisation is necessary in order to obtain a finite entanglement entropy from any formula; the option provided by the above formulation to implement a covariant spacetime ultraviolet cutoff (such as one in terms of a minimum spacetime volume) is physically desirable in gravitational settings. We will make use of this in our example in Section \ref{sec: ee example}.

 As will become more evident in the example calculation in Section \ref{sec: ee example}, the direct appearance of the spacetime domain in e.g. \eqref{eq: gen eig boson explicit}, through the arguments of $W(x,x')$ and $i \Delta(x,x')$, and the domain of the $L^2$ inner product, readily lends itself to calculations of entanglement entropy in arbitrary and nontrivial spacetime regions. This is because adapting the problem to such regions merely amounts to a restriction of the domains to these regions. Similarly, the explicit appearance of the spacetime domain in this formulation also facilitates discrete  numerical calculations, as working with a discrete spacetime, be it fundamental (e.g. a causal set) or not (e.g. a discretisation of the continuum), again becomes a straightforward restriction of domains. This ease to numerically study entropy has the practical benefit of making some otherwise difficult calculations tractable. 

\section{Fermionic Entropy} \label{fermsec}
We will now carry out a similar derivation as in the previous section, this time for a Majorana fermion field. Once again writing the density matrix in diagonal form, we have a classical probability distribution over all of the excited states for each mode, with the entropy of the $j^{\text{th}}$ mode being 
\begin{equation}\label{eq: fermion mode entropy}
    S_{j} = -\sum_{k_{j}=0}^{1} p_{j,k_{j}} \ln{p_{j,k_{j}}}, \qquad p_{j,k_{j}} = \frac{e^{- \tilde{H}_{j j} k_{j}}}{Z_{j}} = \frac{e^{- \tilde{H}_{j j} k_{j}}}{1+e^{-\tilde{H}_{j j}}}.
\end{equation}
The mode expansion of the field is 
\begin{equation}
    \psi(x) = \sum_{j} a^{\dagger}_{j}f^{*}_{j}(x) + a_{j}f_{j}(x),
\end{equation}
where $^*$ denotes charge conjugation, which acts as $^*: \varpsi \to C \bar{\varpsi}^{T}$. $C$ is the charge conjugation matrix, satisfying
\begin{equation}
    C \gamma^{\mu} C^{-1} = -(\gamma^{\mu})^{T}.
\end{equation}

The fermionic Fock space only admits $k_j=0$ or $k_j=1$. We can therefore write the quasifree two-point correlation function $W$ as 
\begin{equation}
\begin{split} \label{eq: fermionic W}
    \langle \Psi|\psi(x) \bar{\psi}(x')|\Psi\rangle &= \text{tr}(\rho\, \psi(x) \bar{\psi}(x')) \\
    &= \sum_{j} p\big(|k\rangle_{j}=|0\rangle_{j}\big) f_{j}(x)\bar{f}_{j}(x') + p\big(|k\rangle_{j}=|1\rangle_{j}\big) f^*_{j}(x)\bar{f}^{*}_{j}(x')\\
    &= \sum_{j} p_{j,0} f_{j}(x)\bar{f}_{j}(x')
    + p_{j,1} f_{j}^{*}(x)\bar{f}^{*}_{j}(x').
\end{split}
\end{equation}
In this case, we can extract the single degree of freedom probabilities as each pair of coefficients in the sum above. Noting that the form of the anticommutator $i \Delta_F$ is 
\begin{equation} \label{eq: anti-commutator}
    \{\psi(x), \bar{\psi}(x')\} = \sum_{j} f_{j}(x)\bar{f}_{j}(x') + f_{j}^{*}(x)\bar{f}^{*}_{j}(x'),
\end{equation}
we see that we can again extract the probabilities necessary for the calculation of entropy via a spectral approach, by promoting \eqref{eq: fermionic W} and \eqref{eq: anti-commutator} to operators via the duality map $\mathscr{A}_{\llangle[\scriptscriptstyle]\cdot,\cdot\rrangle[\scriptscriptstyle]}: f \to \llangle \bar{f}, \cdot \rrangle$, where $\llangle \cdot, \cdot \rrangle$ is the inner product on the one-particle Hilbert space. In this case, one conventionally takes the Dirac inner product, given by
\begin{equation}
    \llangle \varpsi, \tilde\varpsi \rrangle_D := i\int_{\Sigma} \bar{\varpsi}(x)\, \slashed{n}\, \tilde\varpsi(x)\, dS,
\end{equation}
where again $\Sigma$ is a Cauchy surface, with a unit normal vector $n^\mu$,  and $dS$ is the induced volume element on $\Sigma$. Once again, we can free the spectral formulation from the hypersurface, using the fact that $W$ and $\Delta_{F}$ are formed of solutions, and leveraging a similar relation to before \cite{Dimock1982DiracQF},
\begin{equation}\label{eq: relation between D and L2}
    \llangle \varpsi, \tilde\varpsi \rrangle_D = i\int_{\Sigma} \bar{\varpsi}(x)\, \slashed{n}\, \tilde\varpsi(x)\, dS = -i\int_{\mathcal{M}} \bar{\varpsi}(x) h(x)\, dV =: -i\llangle \varpsi, h \rrangle_{L^{2}}.
\end{equation}
Here both $\varpsi$ and $\tilde\varpsi$ are solutions to the equation of motion, and $\tilde\varpsi = \hat\Delta_{F} h$, $h\in C^{\infty}_{0}(D\mathcal M)$.
The generalised eigenvalue problem we then need to solve is
\begin{equation}\label{eq: gen eig ferm}
    \hat{W} h = i\lambda\hat{\Delta}_F h.
\end{equation}
In \eqref{eq: gen eig ferm}, we have defined the operators 
\begin{align}
    (\hat{W}f)(x) &:= \langle\Psi|\psi\, \mathscr{A}_{L^{2}} (\bar{\psi})|\Psi\rangle f,\label{eq: WF as operator} \\
    (i \hat{\Delta}_F f)(x) &:= \{\psi,\, \mathscr{A}_{L^{2}} (\bar{\psi})\} f,\label{eq: DeltaF as operator}
\end{align}
where as in the bosonic case, $\mathscr{A}_{L^{2}}$ is the duality map induced by the $L^{2}$ inner product, $\mathscr{A}_{L^{2}}:f \to \llangle \bar{f}, \cdot \rrangle_{L^{2}}$, which acts on the spinor space, but not the algebraic part. In \eqref{eq: WF as operator} and \eqref{eq: DeltaF as operator}, $f$ is arbitrary. As we have chosen to express our eigenvalue equation in spacetime, the space we must solve over is $h \in C^{\infty}_{0}(D\mathcal M)/\operatorname{ker}(\hat{\Delta}_{F})$, as we must ``mod out'' the kernel of $\hat{\Delta}_{F}$ for the same reasons as in the bosonic case. Much like with the bosonic case, given all conventional choices, \eqref{eq: gen eig ferm} is
    \begin{equation}
        \int_{\mathcal{M}} W(x,x') h(x')\,dV' = i\lambda\int_{\mathcal{M}}\Delta_{F}(x,x') h(x')\,dV'.
    \end{equation}
The eigenfunctions $h(x)$, of \eqref{eq: gen eig ferm}, are such that\footnote{Our $h$'s are again representatives of equivalence classes, that map to a single $f$ or $f^{*}$ under the action of $\hat{\Delta}_{F}$.}
\begin{align}
    \int_{M}\Delta_{F}(x,x')\,h^{A}_{j}(x')\, dV' &= f_{j}(x), \\
    \int_{M}\Delta_{F}(x,x')\,h^{B}_{j}(x')\, dV' &= f^{*}_{j}(x),
\end{align}
and they have corresponding eigenvalues
\begin{align}
    \lambda^{A}_{j} &= p_{j,0} = \frac{1}{1+e^{-\tilde{H}_{j j}}},\label{eq: pj0 fermion} \\
    \lambda^{B}_{j} &=  p_{j,1} = \frac{e^{- \tilde{H}_{j j}}}{1+e^{-\tilde{H}_{j j}}} = 1-\lambda^{A}_{j}.\label{eq: pj1 fermion}
\end{align}
The eigenvalues satisfy $\frac{1}{2}\leq \lambda^{A}_{j}\leq 1$ and $0\leq \lambda^{B}_{j}\leq \frac{1}{2}$. We can then obtain the full entropy by inserting the probabilities, which in this case are the $\lambda$'s, into the sum
\begin{equation}
    S = -\sum_{\lambda} \lambda \ln \lambda \label{Ssumf}.
\end{equation}
In Appendix \ref{appendix: algebraic S for fermions} we derive \eqref{eq: gen eig ferm} and \eqref{Ssumf} in a more algebraic manner.

It is interesting to note the close similarity between the bosonic and fermionic expressions for the generalised eigenvalue equation, \eqref{Sev} and \eqref{eq: gen eig ferm}. The main difference is that in the bosonic theory the right hand side of \eqref{Sev} uses the commutator $i\Delta$, while the fermionic expression \eqref{eq: gen eig ferm} contains the anticommutator  $i\Delta_F$. This difference is intuitive due to the analogous roles of the commutator and anticommutator in the spin statistics of these theories. The entropy as a sum of the solution spectrum of the generalised eigenvalue equation, \eqref{Ssumf}, is also of the same form as its bosonic counterpart \eqref{Ssum} (but with an overall minus sign). There is no absolute value in the argument of the logarithm in \eqref{Ssumf}, as all the $\lambda$'s are nonnegative. For pure states we  will have $\lambda=0$ or $\lambda=1$, resulting in a vanishing entropy from \eqref{Ssumf}, as expected.

This spectral and spacetime formulation of the fermionic entropy also has the same benefits that were discussed at the end of the previous section. Note that although the above analysis was conducted for the case of Majorana fermions, any fermion field can be decomposed into Majorana degrees of freedom. Each Majorana degree of freedom in such a decomposition can be treated independently, as the details of how they combine are not relevant to the entropy calculation.

\section{An Entanglement Entropy Calculation in a Discrete Setting}\label{sec: ee example}
We now turn our attention to an example which uses  several of the strengths of our entropy formulation, namely, the application to a discrete setting with no notion of a Cauchy surface, the use of a covariant cutoff, and the convenience of the generalised eigenvalue equation in numerical calculations and when the density matrix is not already diagonalised. The example is that of the entanglement entropy of a massless scalar field on a causal set approximated by a Rindler wedge in $1+1$ dimensions. For a brief demonstration of how the entropy formulae reproduce known results in a simple thermal case, we refer the reader to Appendix \ref{examplesec}. We begin with a review of causal set theory in Section \ref{sec: cst}, describe the spacetime background and quantum field setup of our calculation in Sections \ref{sec: ee setup} and \ref{subsec: SJ state in diamond} respectively, and summarise the results in Section \ref{sec: ee results}.
\subsection{Causal Set Theory}\label{sec: cst}
Causal set theory is an approach to quantum gravity that makes minimal assumptions, while incorporating key physical principles. It posits the prominence of causal structure, as well as a fundamental discreteness of spacetime that respects general covariance. 

The requirements of discreteness, and the macroscopic recovery of spacetime as we know it from general relativity, seem to necessitate an irregular, spacetime volume-based, discrete structure known as a causal set \cite{ButterfieldDowker2024}. This causal set must encode  causal structure, and so should be a directed acyclic graph, or equivalently, a partially ordered set of spacetime elements \cite{BLMS1967}. Each element of the causal set has associated to it (statistically) a minimum spacetime volume, $V_{\text{min}}$, which is the smallest physically meaningful volume. It is preferable to work with minimum volume scales, as they are inherently covariant, but one can also view this minimum volume as inducing a minimum length scale of $l_{\text{min}} = \sqrt[d]{V_{\text{min}}}$.

Causal set theory then proposes that the fundamental structure underlying spacetime is a causal set. While replacing the smooth Lorentzian manifold and metric of general relativity with a causal set is motivated by quantum gravity, there are already consequences and insights that arise at the classical and semiclassical levels. For example, classically, we learn how familiar smooth geometric quantities such as timelike and spacelike distances \cite{PhysRevLett.66.260, Boguna:2024jye} as well as differential operators such as the d'Alembertian \cite{Sorkin:2007qi, Benincasa:2010ac} can emerge from discrete (often combinatorial) quantities in a causal set. Semiclassically, placing quantum fields on a background causal set UV regulates quantities such as the Wightman function. However, the discrete, spatiotemporal, and frame independent nature of causal sets gives rise to nonlocality, and precludes Cauchy surfaces from existing in them.\footnote{See footnote \ref{ft: no Cauchy} for more details.} Therefore, conventional canonical quantisation techniques do not apply. It was in fact this need for a quantisation prescription that does not rely on a Cauchy surface that led to the construction of the Sorkin-Johnston (SJ) state, which we review in Appendix \ref{ap: SJ} and which we use in this section. The SJ state has several other desirable properties, such as being frame independent and pure, and is well defined in any spacetime (discrete or continuum) with a retarded Green function.\footnote{The SJ state is unique if the spacetime possesses unique retarded and advanced Green functions (e.g. in globally hyperbolic spacetimes).} It is also worth mentioning that the entropy formulation discussed in this work was originally motivated by the desire to have a spacetime definition of entropy that could be used in a causal set (in addition to the continuum), and in particular, able to make use of its intrinsic covariant volume-based cutoff.

We will henceforth take a classical causal set as our background, and define a quantum real scalar field on it using the Wightman function in the SJ state, as we explain more in the next subsection. The standard practical procedure to generate a causal set approximated by (or approximating) a given spacetime, is by \emph{sprinkling}: one starts with the smooth manifold and samples points from it via a Poisson process.\footnote{The manifold is not in principle needed to generate a ``manifoldlike" causal set. However, at this stage of the development of causal set theory, using a manifold to generate a manifoldlike causal set is most convenient.}  After sprinkling, the probability of there being $N$ points in a region with spacetime volume $V$ is given by the Poisson distribution, 
\begin{equation}
P_N(V) = \frac{(\tilde{\rho} V)^N}{N!}e^{-\tilde{\rho} V}, 
\end{equation}
with mean $\langle N\rangle=\tilde{\rho} V$ and standard deviation $\Delta N=\sqrt{\tilde{\rho} V}$. This has been shown to produce the best ``number-volume" correspondence \cite{NVcorr}, which is necessary for the causal set to contain the volume information of the approximating spacetime.  The sprinkled elements are then given causal relations according to the underlying spacetime, and in this way, together with the volume information, we capture all the information of the spacetime above the discreteness scale \cite{Zeeman, Hawking:1976fe, Malament, Braun:2025enn}. The resultant sprinkled causal set is characterised by the constant density, $\tilde{\rho}=\frac{\langle N\rangle}{V}$, and the minimum volume scale is $V_{\text{min}} := \frac{1}{\tilde{\rho}}$. 

We refer the reader to \cite{SuryaReview} for an overview of other aspects of causal set theory that are not directly related to our work. 
\subsection{Entanglement Entropy in a Rindler Wedge Causal Set}
\subsubsection{Spacetime Setup}\label{sec: ee setup}
\begin{figure}
    \centering
    \begin{subfigure}[b]{0.49\textwidth}
        \centering
        \includegraphics[width=\textwidth]{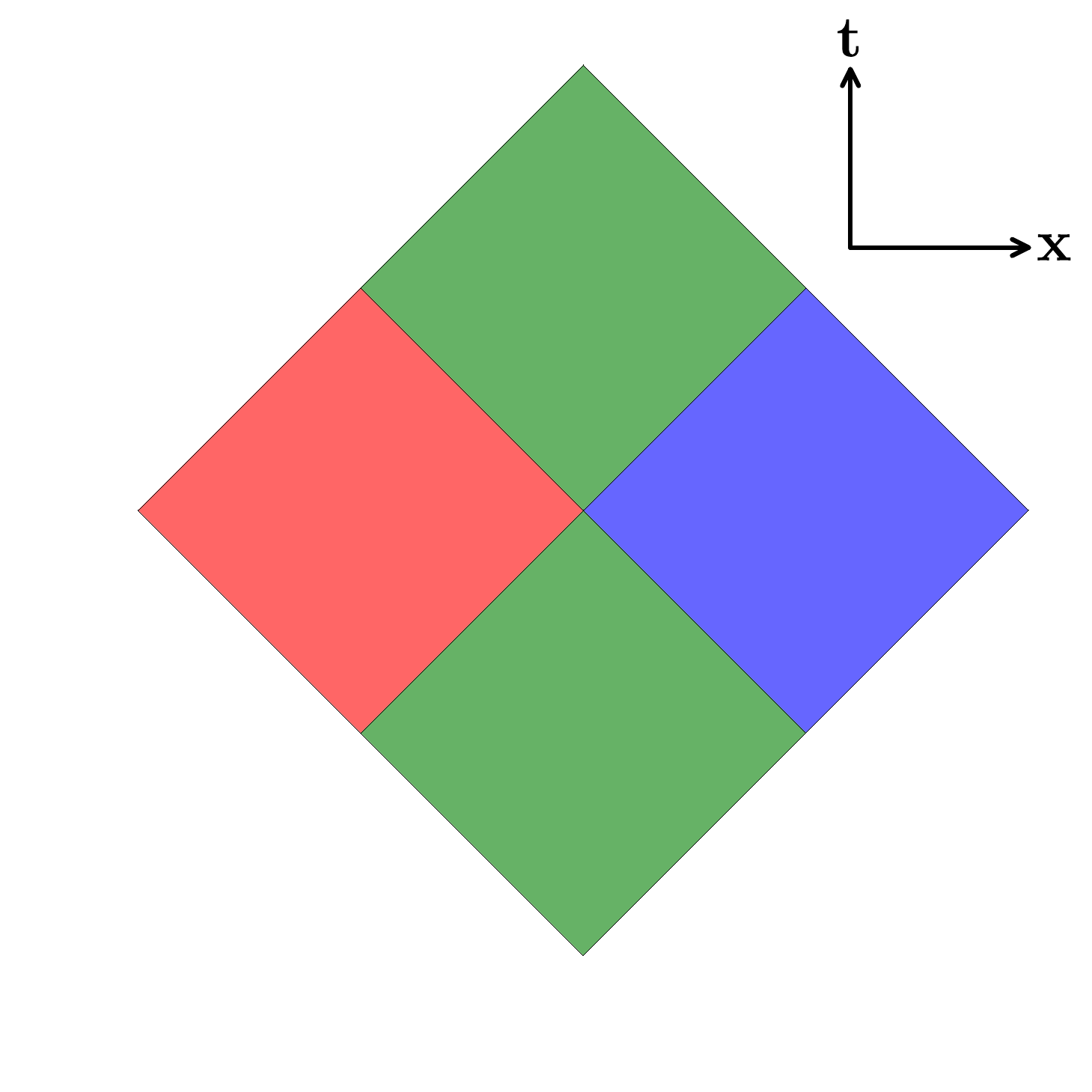}
        \caption{}
        \label{fig:Setup}
    \end{subfigure}
    \hfill
    \begin{subfigure}[b]{0.49\textwidth}
        \centering
        \includegraphics[width=\textwidth]{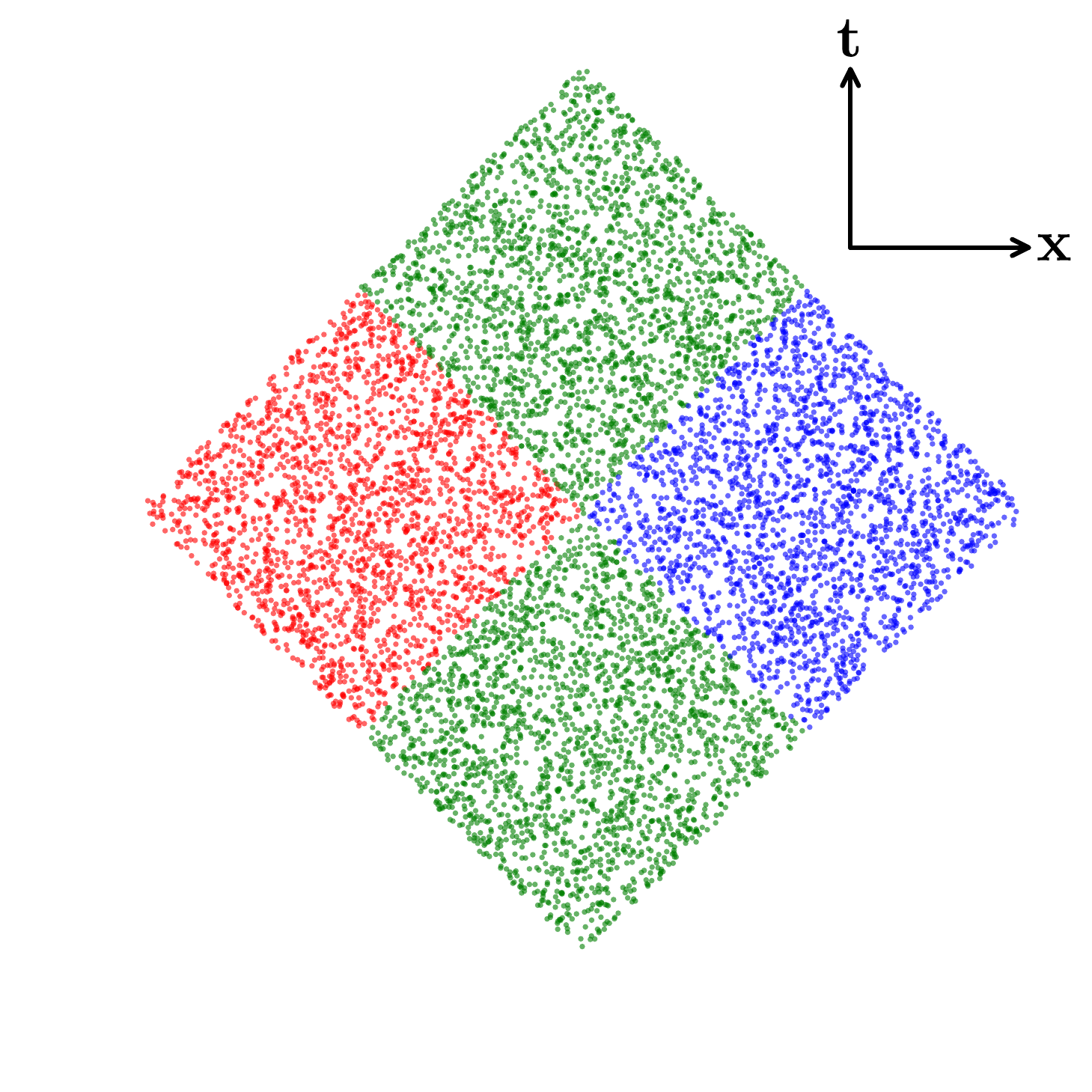}
        \caption{}
        \label{fig:CST_Setup}
    \end{subfigure}
    \caption{(a) The full spacetime under consideration, with the left subdiamond in red, and the right subdiamond in blue. The red and blue subdiamonds are Rindler wedges with an infrared cutoff, and they are causal complements of one another. (b) A sprinkling of the spacetime on the left. The points sprinkled into the left and right subdiamonds are shown in red and blue respectively; we will restrict $W$ and $i\Delta$ to either the red or the blue elements when we calculate the entanglement entropy via \eqref{Sev} and \eqref{Ssum}.}\label{fig: diamond setup}
\end{figure}

We will consider a massless scalar field on a Rindler wedge in $1+1$d Minkowski spacetime with an infrared cutoff, which is needed to tame the infrared divergence of the massless theory in $1+1$d. We implement this cutoff by taking our full spacetime to be a large causal diamond with side length $2L$ (in $1+1$d Minkowski spacetime), centred at the origin of the $\{t,\mathbf{x}\}$ plane, instead of infinite Minkowski spacetime. This diamond is  then quadrisected into four smaller subdiamonds (each with side length $L$), as shown in Figure \ref{fig:Setup}. The left (red) and right (blue) subdiamonds correspond to the left and right wedges of Rindler spacetime (with an infrared cutoff). This setup also facilitates the causal set calculation, as only a finite number of elements is needed to sprinkle the diamond. The sprinkling is shown in Figure \ref{fig:CST_Setup}. We then calculate the entanglement entropy resulting from  the field being restricted to either the left or right subdiamond. In the causal set context, this amounts to calculating the entanglement entropy of the field restricted to either the corresponding red or blue elements in Figure \ref{fig:CST_Setup}, which will produce a nonzero entropy, as the causal complement of the  elements (region)  we restrict to is non-empty.

A similar calculation to the one in this section appears in \cite{SSY2014} for the continuum and in \cite{SY2018} for the analogous causal set setup. There the entanglement entropy considered was the result of restricting the field to a subdiamond concentric with the full spacetime causal diamond. In \cite{SSY2014} the additional simplification was made that the size of the subdiamond was set to be much smaller than the size of the larger diamond. Other related calculations include \cite{DJY2022, KMY2022, SNY2021, MSN2022}. Here we have chosen the setup of Figure \ref{fig: diamond setup} due to its relation to Rindler spacetime which is a quintessential spacetime for the study of entanglement entropy. 

 The retarded and advanced Green functions for the massless theory in $1+1$d have simple expressions in the causal set. These Green functions can be obtained either via a path integral style construction, or by direct identification with the causal set adjacency matrix as we explain below. These can then be used to construct $i\Delta$, and subsequently $W_{SJ}$, the Wightman function in the SJ state, which is the state conventionally chosen in the causal set \cite{JohnstonThesis, Sorkin2011, DiamondGS}. With $i\Delta$ and $W_{SJ}$ at hand, the entropy can be computed via \eqref{Sev} and \eqref{Ssum}. The SJ state in a causal diamond resembles the Minkowski vacuum away from the boundaries of the diamond. Boundary effects exist near the left and right corners of the diamond, wherein the state resembles the Minkowski vacuum with mirrors (reflecting boundaries) at those corners \cite{DiamondGS}. 

\subsubsection{$W_{SJ}$ in a Causal Diamond}\label{subsec: SJ state in diamond}
\paragraph{Continuum Theory:}
We begin with a discussion of the functions and operators that we will need, in their continuum form. The retarded and advanced Green functions for a massless scalar field in $1+1$d are

\begin{equation}\label{eq: gret massless 2d}
    G_R(x,x')=\frac 1 2 \Theta(t-t') \Theta(\tau^2(x,x'))=G_A^T(x,x'),
\end{equation}
where $\Theta$ is the Heaviside step function and $\tau(x,x')$ is the proper time between spacetime points $x$ and $x'$.

It is convenient to work in lightcone coordinates, where $u = \frac 1 {\sqrt{2}} (t+\mathbf{x})$ and $v = \frac 1 {\sqrt{2}} (t-\mathbf{x})$. We obtain $i\Delta(x,x')$ by substituting \eqref{eq: gret massless 2d} into \eqref{eq: iDelta commutator}, and promote it to an integral operator using the $L^2$ inner product. We can find the eigendecomposition of the Pauli-Jordan function by solving
\begin{equation}\label{eq: PJeigen}
    i\hat{\Delta}f = \tilde \lambda f.
\end{equation}
In our causal diamond, \eqref{eq: PJeigen} takes the form 
\begin{align}
\label{eq: PJeigen diamond}
    (i\hat\Delta f)(u,v) &= \frac{i}{2} \int_{\mathcal{M}} (\Theta(u-u') + \Theta(v-v') - 1) f(u',v') \,dV',\\
   & = \frac{i}{2} \int_{-L}^L\int_{-L}^L (\Theta(u-u') + \Theta(v-v') - 1) f(u',v') \,du'\, dv',
\end{align}
where in the second line the region of integration  $\mathcal{M}$ was set to be our causal diamond of side length $2L$, centered at the origin. The two sets of solutions of  (\ref{eq: PJeigen}) are \cite{JohnstonThesis}
\begin{equation}\label{eq: diamondeigs1}
    f_k(u, v) = e^{-iku} - e^{-ikv}; \qquad k = \frac{n\pi}{L}, \quad n \in \mathbb{Z}\backslash\{0\},
\end{equation}
and
\begin{equation}\label{eq: diamondeigs2}
    g_k(u, v) = 
    e^{-iku} + e^{-ikv} - 2\cos{(kL)}; \qquad k \in \mathcal{K}= \{k \in \mathbb{R} | \tan{(kL)}=2kL \text{ and } k \neq 0 \},
\end{equation}
each with an eigenvalue
\begin{equation}\label{eq: diamond eigenvalue L over k}
    \tilde \lambda_{f_k} = \frac{L}{k},\qquad \tilde \lambda_{g_k} = \frac{L}{k}, 
\end{equation}
with their corresponding values of $k$ as defined in \eqref{eq: diamondeigs1} and \eqref{eq: diamondeigs2}. We can hence see that the eigenfunctions are a linear combination of two plane waves with the same wavenumber (plus a constant contribution for  $g_k$). We also see from the equations above that the characteristic length scale of the eigenfunctions (or modes), $k$, is encoded in the magnitude of their eigenvalue. In other words, the more ultraviolet the scale of variation of an eigenfunction is, the smaller the magnitude of the eigenvalue. This is crucial for understanding the ultraviolet cutoff, for example, for the purpose of obtaining a finite answer in entanglement entropy calculations. One way to implement this cutoff is to express $i\Delta$ and $W$ in terms of a finite set of eigenfunctions up to a $k_{max}$ with the value of $k_{max}$ serving as the UV cutoff. This was done in \cite{SSY2014}.

Finally, we also note that for large $k$, the eigenvalues of the $g_k$ eigenfunctions tend to those of the $f_k$ eigenfunctions, which have the form
\begin{equation}\label{eq: spectral density}
  \tilde  \lambda_{f_k} = \frac{L^2}{n\pi}.
\end{equation}

Using the eigenvalues and eigenfunctions above, the SJ Wightman function, i.e. the positive spectral part of $i\Delta$ ($W_{SJ} = \text{pos}(i\Delta)$), can be constructed as 
\begin{equation}
\begin{aligned}
    W_{SJ}(u,v;u',v')=&\sum_{f_k}\frac{\Theta(\tilde \lambda_{{f_k}})\tilde\lambda_{{f_k}}}{\llangle f_k, f_k \rrangle_{L^2}} f_k(u,v) f^*_k(u',v') \\
    +&\sum_{g_k}\frac{\Theta(\tilde\lambda_{{g_k}})\tilde\lambda_{{g_k}}}{\llangle g_k, g_k \rrangle_{L^2}} g_k(u,v) g^*_k(u',v').
\end{aligned}
\end{equation}

\paragraph{Causal Set Theory:}
We now turn to the causal set analogue of the above construction. The retarded Green function \eqref{eq: gret massless 2d} has an exact counterpart\footnote{As mentioned earlier, the causal set retarded Green function in this case can be interpreted in two ways: (1) By pragmatically recognising that half the transpose of the causal matrix \eqref{eq: causal matrix} is the direct equivalent to the step function on the past lightcone, which constitutes the continuum version of this function, or (2) As a path integral (or sum) over chains between pairs of elements of the causal set, as described in Chapter 3 of \cite{JohnstonThesis}.}: 
\begin{equation}\label{eq: gret massless 2d cst}
    G_R=\frac{1}{2} C^{T},
\end{equation}
 where $C$ is the causal matrix defined as
\begin{equation}\label{eq: causal matrix}
    C_{xx'}:=\begin{cases}
    1, & \text{if $x\prec x'$}\\
    0, & \text{otherwise},
    \end{cases}
\end{equation}
where the subscript $xx'$ refers to the matrix entry corresponding to elements $x$ and $x'$, and $x\prec x'$ means that $x$ causally precedes $x'$. We then proceed  as above. We obtain $i\Delta$ using \eqref{eq: gret massless 2d cst} in \eqref{eq: iDelta commutator}, find its ($L^2$-normalised) eigenvectors $\{v_i\}$ and eigenvalues $\{\tilde\lambda_i\}$,\footnote{Note that in the causal set, due to the matrix nature of the relevant objects, we leave the hats off operators.} and then construct the SJ Wightman function as 

\begin{equation}
 W_{SJ} = \sum_{i} \Theta(\tilde\lambda_i) \tilde\lambda_i v_{i} v^{\dagger}_{i}.
\end{equation}

Note that finding the eigenvalues and eigenfunctions of $i\Delta$ is much easier in the causal set context, as here it is a finite-dimensional\footnote{The size is set by the number of elements in the causal set.} matrix rather than the infinite-dimensional integral operator it was in the continuum. There are standard techniques and computational packages that exist for obtaining the eigendecomposition for matrices, and this gives the causal set calculation a significant practical advantage over the continuum, where the corresponding integral equations are difficult to solve.

The eigenvalues $\tilde\lambda_i$ are dimensionless, in contrast to their continuum counterparts \eqref{eq: diamond eigenvalue L over k} which have dimensions of length-squared. We can understand this difference in the following way: if we take our eigenfunction integral equation (\eqref{eq: PJeigen} and \eqref{eq: PJeigen diamond}) to be discretised onto a causal set, we would need to ascribe to the integrand a volume measure, or weight, corresponding to the volume of each element. On average, these weights should come out to be $V_{min}$, or the inverse of the sprinkling density $\frac 1 {\tilde\rho}$, which thus provides us with the conversion factor between the continuum and causal set Pauli-Jordan function spectra.\footnote{\label{conversion between cs and continuum eigenvalues}i.e., for an eigenfunction $f$, $\tilde \lambda_{f} f(x) = \int i\Delta(x,x') f(x') dV' \sim \sum_{x'} i\Delta_{xx'} f_{x'} /{\tilde \rho} = \frac{\tilde \lambda} {\tilde \rho} f_x\implies \tilde \lambda_{f} \approx \tilde \lambda/\tilde\rho$.}

In the causal set, we expect to obtain a set of eigenvectors approximated by the plane wave combinations in \eqref{eq: diamondeigs1} and \eqref{eq: diamondeigs2} up to a $k_{max}$ related to the minimum volume scale, $V_{min}$, with a corresponding minimum eigenvalue $\frac{L}{k_{max}}=\frac{L\, \sqrt{V_{min}}}{2\pi}$. Conversely, we would not expect to have meaningful quantities with variations below this approximate scale. However, upon inspection of the eigendecomposition of $i\Delta$ we see that there are contributions below this scale. This introduces additional subtleties into the problem which we discuss in the next subsection.

\subsubsection{Spectrum Truncation}\label{sec: spectrum truncation}
There is nothing in between the causal set elements, and therefore there do not exist meaningful quantum field degrees of freedom at length scales below the discreteness scale. This leads to an everywhere (including on the diagonal) finite Wightman function. This also regulates the ultraviolet divergence associated with the entanglement entropy, as it removes the responsible (arbitrarily) high frequency modes.  This was a driving motivation for our replacement of the smooth spacetime manifold with a causal set. 

It is indeed the case that on applying the prescription laid out above, we obtain finite quantities, including a finite entanglement entropy. However, there are additional subtleties we need to take into account. These subtleties arise because upon inspection of the spectrum of $i\Delta$ in a causal set diamond, we see that there are a finite (and large) number of residual contributions with magnitudes we would associate to degrees of freedom below the discreteness scale.

The extra contributions below the discreteness scale have an entirely different character compared to the contributions above this scale. The eigenvalue scaling associated with these degrees of freedom is no longer the inverse relation in \eqref{eq: spectral density}, and the eigenfunctions no longer resemble linear combinations of two plane waves. This is illustrated in Figure \ref{fig: loglogeigenvalues} in Appendix \ref{ap: volume law}. It is thus consistent with our intention of using the causal set as a background regulator of ordinary quantum field degrees of freedom, to remove these additional terms. It is likely that these contributions are due to fluctuations tied to the single sprinkling realisations we work with.\footnote{The sprinkling procedure produces a causal set with a uniform and random distribution of elements over a given continuum spacetime region. While the average number of elements approximated by a fixed volume is constant, there will be fluctuations (including below the statistical discreteness scale). The size of these fluctuations is given by the Poisson distribution. These fluctuations also result in the kernel of $i\Delta$ being very small in the causal set, as it is rare to get an eigenvalue that is exactly zero.} We would then expect these effects to go away when some suitable averaging of the eigenfunctions is done over different sprinklings (c.f. \cite{KMY2022}), or else when the calculation is set in a framework beyond the semiclassical theory, and in particular, without a fixed background. Nevertheless, for the case at hand, there is a straightforward procedure to modify our analysis to take these subtleties into account. We describe this \emph{spectrum truncation} procedure below.

 The average minimum volume $V_{min}$ induces a minimum length $\frac{1}{\sqrt{\tilde\rho}}$, which we expect to be the approximate minimum wavelength described by our eigenfunctions. Associated to this scale, we have the corresponding maximum wavenumber and minimum magnitude (continuum) eigenvalue of $i\Delta$,
\begin{equation}\label{eq: min eig cont}
    k_{\text{max}} = 2\pi \sqrt{\tilde \rho} \iff |\tilde \lambda_{\text{\tiny min}}^{\text{\tiny cont}}| = \frac{L}{2 \pi \sqrt{\tilde\rho}}.
\end{equation}
We then multiply by $\tilde \rho=\frac{\langle N_{2L}\rangle}{4L^2}$ (see footnote \ref{conversion between cs and continuum eigenvalues}) to find the dimensionless causal set minimum magnitude eigenvalue 
\begin{equation}\label{eq: min eig cs}
    |\tilde\lambda_{\text{\tiny min}}^{\text{\tiny cs}}|\sim\tilde \rho\, |\tilde \lambda_{\text{\tiny min}}^{\text{\tiny cont}}| = \frac{\sqrt{\langle N_{\scriptscriptstyle{2L}}\rangle}}{4\pi},
\end{equation}
where $\langle N_{\scriptscriptstyle{2L}}\rangle$ is the mean cardinality of the causal set diamond we are considering. The subscript of $N$ refers to the side length of the diamond under consideration. We can hence conclude that if the degrees of freedom below the discreteness scale of the causal set are absent, we expect no nonzero eigenvalues of $i\Delta$ with magnitude smaller than approximately $\tilde\lambda_{\text{\tiny min}}^{\text{\tiny cs}}$. Since this is not automatically enforced by the causal set, we must remove (e.g. via a projection) any eigenvectors with  eigenvalue smaller in magnitude than  \eqref{eq: min eig cs}. This procedure is often called a \emph{truncation}, and must be done for both $W_{SJ}$ and $i\Delta$. Note that the relation $W_{SJ} = \text{pos}(i\Delta)$ is preserved under this truncation.

Post truncation, we have spacetime operators that have the unwanted residual short distance degrees of freedom stripped out. We then make a restriction to either the red or blue subdiamond in Figure \ref{fig:CST_Setup}. After restriction, we again need to truncate both $W_{SJ}$ and $i\Delta$. The spectra of the truncated operators after restriction is much like that of the initial operators,\footnote{ There is a regime of large (magnitude) eigenvalues following a power law with corresponding eigenfunctions which are continuumlike, after which there is a non-power law tail of small eigenvalues corresponding to non-continuumlike eigenfunctions.} and therefore this second truncation serves to once again remove any degrees of freedom with a characteristic wavelength shorter than the discreteness scale. The exact same reasoning applies in the case of this second truncation, and so we express $i\Delta$ and $W_{SJ}$ in the eigenbasis of $i\Delta$ in the subdiamond and project out the parts associated with eigenvectors with an  eigenvalue smaller in magnitude than
\begin{equation}
    {|\tilde \lambda_{\text{\tiny min}}^{\prime\,\text{\tiny cs}}}| \sim \frac{\sqrt{\langle N_{\scriptscriptstyle{L}}\rangle }}{4\pi},
\end{equation}
where $\langle N_{\scriptscriptstyle{L}}\rangle$ is the mean cardinality of the causal set subdiamond we restricted to. This spectrum truncation also serves to  remove the kernel of $i\Delta$, which is necessary for the entropy calculation. For further discussion of this truncation, variations of it, and open questions around it, we refer the reader to \cite{SY2018, BBLL2017, Yazdi2023}.

If we do not perform this spectrum truncation, the entanglement entropy will generically follow a spacetime volume law, as found in previous work (e.g. \cite{SY2018, SNY2021}). In Appendix \ref{ap: volume law} we briefly discuss this from the perspective of the equation of motion.

\subsubsection{Results}\label{sec: ee results}
We are now in a position to calculate the entanglement entropy for our setup. We set the side length of our large diamond to be $2L=1$, sprinkle it at a chosen density $\tilde\rho$, calculate the causal matrix \eqref{eq: causal matrix}, and through it obtain the retarded Green function \eqref{eq: gret massless 2d cst} and $i\Delta$. We then diagonalise $i\Delta$ and take its positive part to define the Sorkin-Johnston Wightman function $W_{SJ}$. Finally we carry out the spectrum truncation,  restriction, and second spectrum truncation, as detailed above, before applying our entropy formula. As the $W$ and $i\Delta$ that we insert into \eqref{Sev} are finite-dimensional matrices with numerical entries, we can make use of standard techniques for solving generalised eigenvalue problems.\footnote{We use Arnoldi iteration, since $W$ and $i\Delta$ are sparse  matrices, as they will be in most physical spacetimes.}

We are interested in the scaling of the entanglement entropy with the UV cutoff $\frac 1 {\sqrt{\tilde\rho}}$. We therefore calculate the entanglement entropy for different values of $\tilde\rho$. We vary $\tilde\rho$ by holding fixed the volume of the causal diamond and changing the mean number of elements sprinkled into it, $\langle N_{\scriptscriptstyle{2L}}\rangle$. Since there are fluctuations in the causal set, we moreover calculate the entropy for multiple causal sets at each density $\tilde\rho$, in order to get the mean and standard deviation of the entropy for a given density. Our simulations lie in the range between $\langle N_{\scriptscriptstyle{2L}}\rangle=400$ to $\langle N_{\scriptscriptstyle{2L}}\rangle=102400$, increasing in steps of $\sqrt{\langle N_{\scriptscriptstyle{2L}}\rangle}=20$. At each value of $\langle N_{\scriptscriptstyle{2L}}\rangle$, we perform $30$ simulations, yielding $60$ data points (one from the left subdiamond and another from the right subdiamond). It is also worth highlighting that the results reported here were obtained using the largest sample set and the highest cardinality causal sets out of any other similar calculation on  this subject to date. Our data set and results therefore have a higher precision compared to previous ones.

Our spacetime configuration, Figure \ref{fig: diamond setup}, can be viewed as the domain of dependence of a $t=0$ hypersurface segment. The left or right subdiamond can then be viewed as the domain of dependence of this line segment restricted to $\mathbf{x}<0$ or $\mathbf{x}>0$ respectively. The expected scaling\footnote{In the asymptotic limit of the UV cutoff (in dimensions of length) going to zero, i.e. $a\rightarrow0$.} of the entanglement entropy of a massless scalar field with the UV cutoff, $a$, in such a continuum problem is \cite{CC2004, CC2009}
\begin{equation}\label{eq: ee scaling reference}
    S \sim \frac{1}{6}\log\left(\frac{\tilde{L}}{a}\right) + c,
\end{equation}
where $\frac{\tilde{L}}{a}$ is the interval length in units of the ultraviolet cutoff, and $c$ is a nonuniversal constant that depends, for example, on the implementation of the ultraviolet cutoff. We will compare our results to \eqref{eq: ee scaling reference}.

\begin {figure}[h!]
    \centering
    \includegraphics[width = 15 cm]{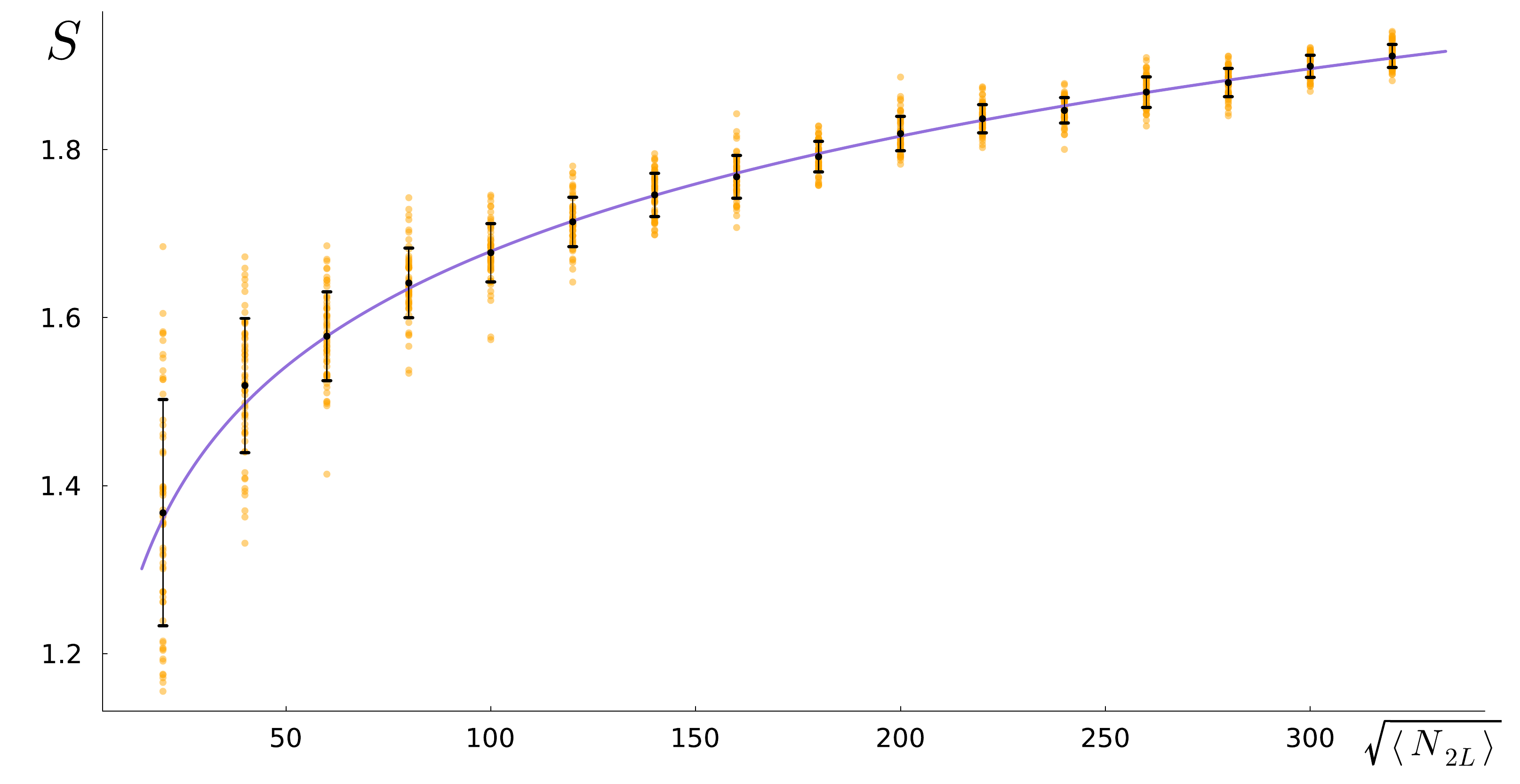}

    \caption {A plot of the results of our simulations. The raw data points are in orange, and the mean and standard deviations of the bins are in black. We have fitted to the data a curve of the form $S=b \log\left(\sqrt{\langle N_{\scriptscriptstyle{2L}}\rangle}\right)+c$, shown in purple, via a least squares regression weighted by the inverse variance.}
\label{fig: Results}
\end {figure}

The results of our simulations are shown in Figure \ref{fig: Results}. We have plotted the entropy versus $\sqrt{\langle N_{\scriptscriptstyle{2L}}\rangle}$, which is proportional to the inverse of the cutoff $\sqrt{\tilde\rho}=\frac{\sqrt{\langle N_{\scriptscriptstyle{2L}}\rangle}}{2L}$. The raw data are shown in orange. The mean and standard deviation for each ensemble of causal sets at fixed $\tilde\rho$ are shown by the black circle and vertical error bar line respectively. We fit  the function $b \log\left(\sqrt{\langle N_{\scriptscriptstyle{2L}}\rangle}\right)+c$ to our data, and we obtain the values $b=0.198\pm0.014$ and $c=0.767\pm0.076$. This best fit is shown by the purple curve in Figure \ref{fig: Results}. The fit and uncertainties were calculated using a least squares regression, weighted by the inverse of each bin's variance. 

We see that our results follow a logarithmic scaling with the cutoff, as expected, although with a coefficient that is slightly larger than the continuum value of $\frac 1 6$. This discrepancy does not appear to get smaller as we increase the number of elements in the causal set and therefore seems to be a genuine difference between the continuum and causal set. Before further discussing this coefficient, let us revisit how a logarithmic scaling in $1+1$d fits in with entanglement entropy area laws, following the argument in \cite{SorkinAreaLaw}. Consider a spatial region $\Sigma$ which is a Cauchy surface, and a field (initially in its ground state) whose entanglement entropy we wish to study when restricted to a subregion $\Sigma^\prime$ of $\Sigma$. If we imagine the field to be composed of approximately localised modes, then the entropy will be dominated by (roughly equal) contributions from the modes straddling the boundary of $\Sigma^\prime$. To find the scaling properties of the entropy, we must then count the number of these straddling modes, i.e. $S \propto N_{strad}$. 

This picture also shows why the entropy would be infinite without a UV cutoff, as in such a case there would be an infinite number of modes  straddling the boundary. We will set a UV cutoff $k_{max}$, and an IR cutoff (for a massless theory) $k_{min}$. In the neighbourhood $d^{d-1}k$ of wave vector $\vec{k}$, the number of modes astraddle $\partial\Sigma'$ will be approximately $V_{\Sigma^\prime} \frac{d^{d-1}k}{(2\pi)^{d-1}}$, where $V_{\Sigma^\prime}=\frac{2\pi A_{\partial\Sigma'}}{|\vec{k}|}$, and $A_{\partial\Sigma^\prime}$ is the area of $\partial\Sigma^\prime$. Integrating this, we obtain the total number of straddling modes, and thereby the entropy scaling

\begin{equation}\label{eq: area law}
    S \propto N_{strad}\sim\int^{k_{max}}_{k_{min}}\frac{2\pi A_{\partial\Sigma'}}{|\vec{k}|} \frac{d^{d-1}k}{(2\pi)^{d-1}}=\frac{ 2^{3-d}\pi^{\frac{3-d}{2}} }{ \Gamma(\frac{d-1}{2})}\int^{k_{max}}_{k_{min}}   A_{\partial\Sigma'}\, k^{d-3} dk.
\end{equation}

The integral in \eqref{eq: area law} results in an area law for $d>2$, and a log-corrected area law for $d=2$. The log-corrected area law in $d=2$ is proportional to the ``area” $A_{\partial\Sigma'}$ of the boundary of the subinterval, and this boundary consists of one (as in our setup Figure \ref{fig:Setup}) or more points. In our causal set setup (Figure \ref{fig:CST_Setup}), we do not have such a sharp spatial  interval or boundary.\footnote{\label{ft: no Cauchy} A maximal antichain -- the largest set of causally unrelated elements such that every other element is causally related to at least one of the elements of this set -- is the causal set analogue of a spatial slice. However, since there will typically be some causal relations that are not deducible from the relations known to this set, it is not the analogue of a Cauchy surface. We may supplement the maximal antichain with causally related elements to help register the missing relations and bring it closer to being an approximate Cauchy surface. As such a set would contain some causally related elements, it is a kind of \emph{thickened} Cauchy surface. In a setting like our diamond, more thickening would be needed in the middle of the diamond (away from the boundaries) compared to the edges, as this region's lightcones cover a larger volume of the spacetime and would hence be more likely to miss relations.} It is therefore likely that to the extent that we can relate such a picture to our causal set setup, the boundary across which the modes straddle would be slightly blurred, or thickened, rather than consisting of a single element; this would have the effect of increasing $A_{\partial\Sigma'}$ and we can see from \eqref{eq: area law} that it would then increase the prefactor of the logarithm. This may be the reason for the larger coefficient we obtained. For a discussion of what this may look like in terms of a (nonlocally) modified equation of motion and $\Delta$, see Appendix \ref{ap: volume law}. If this interpretation is correct, then it is interesting that there is a small yet nonnegligible signature of the discreteness of spacetime in the scaling law.\footnote{Note that even if a different explanation for this excess entropy scaling turns out to be correct, it should nevertheless be linked to the discreteness of the system, as this excess is absent in similar continuum settings \cite{SSY2014}.}${}^{\;\!\!, \:\!}$\footnote{Another possibility is that we have not truncated enough of the residual non-continuumlike contributions by setting our minimum wavelength \eqref{eq: min eig cont} to be exactly the discreteness scale. These potentially extra contributions could be making up the excess entropy. We repeated our analysis with the minimum wavelength set to twice the discreteness scale and found that while the coefficient of the logarithm fit became closer to $\frac 1 6$, it was still larger ($b=0.182\pm0.007$), making this explanation less likely.}

\section{Conclusion}
We have presented a spacetime definition for the entropy of quasifree (or Gaussian) bosonic and fermionic theories. We derived the formulae for this entropy, showing the relation to the well-known $S = -\text{Tr}\left( \rho_{red} \ln\rho_{red}\right)$ expression, highlighted some key aspects of the prescription, and gave example applications. 
In our presentation, we have taken care to develop  the connection between this definition of entropy and other known formulations in the literature. In particular, the derivations in Sections \ref{bosonsec} and  \ref{fermsec} are in the language of Hilbert spaces, where other formulations of entropy tend to reside. These derivations are entirely new. The algebraic presentation in Appendix \ref{algebrasec} is also a more detailed  exposition of its kind, and the fermionic derivation in that appendix is new. In Appendix \ref{examplesec}, we show how one can use the spacetime entropy definition to calculate the thermal entropy of a scalar and fermionic field in $3+1$d Minkowski spacetime and obtained the expected answers, verifying that the formulae reproduce known results. In Section \ref{sec: ee example}, we carried out a calculation that was only made possible by this formulation, and which put to practice many of its useful features: we computed the entanglement entropy of a scalar field defined on, and regularised by, a discrete spacetime (or causal set) approximated by a Rindler wedge in $1+1$d. The entropy results from this calculation followed a logarithmic scaling in the cutoff (set by the discreteness scale), with a coefficient that was slightly larger than what is obtained from similar continuum calculations. This greater coefficient is consistent with modifications expected from the discreteness.

Physically, the spacetime nature of this entropy formulation provides a favourable framework for questions concerning (quantum) gravity, such as the relationship between entanglement entropy and the microscopic origin of horizon entropy. This definition of entropy enables the use of a covariant spacetime volume-based cutoff (such as the one provided by a causal set) to regularise the otherwise infinite entanglement entropy. We  demonstrated this in our example in Section \ref{sec: ee example}. The ability to work with a covariant cutoff is essential to unambiguously make statements about horizon entropy. Without it, one has to resort to arguments involving preferred frames, which are neither unique nor justified. Moreover, such frame dependent arguments have limited scope, and are not able to treat all settings in which horizon entropy needs to be understood. In contrast, our formulation, and similar calculations to the one in the previous section, can be applied in general spacetimes such as black hole (with different dimensions, mass, etc.) and cosmological (with different energy contents, asymptotics, etc.) spacetimes. This entropy formulation therefore marks a significant step towards understanding the microscopic degrees of freedom that characterise  horizon entropy, as a universal explanation is suggested by the ubiquity of horizon entropy area laws.

There are also practical benefits to this entropy definition. As it is expressed in terms of spacetime correlation functions and therefore explicitly in position space, this streamlines many applications. For example, the entanglement entropy associated to arbitrarily nontrivial spacetime regions can be directly considered by simply restricting the domains of the arguments of the correlators to lie in these regions. Discrete spacetimes such as causal sets can be considered, which in addition to their physical benefits, make the (numerical) calculations  considerably more tractable. In these numerical calculations, for computational reasons, finite regions of spacetime must be considered. For large enough regions, the results should approximate those of an infinite region (if there are no infrared divergences). One can also consider the entanglement entropy of spacetime regions that are not globally hyperbolic, and which do not correspond to a hypersurface entropy at all.

Our formalism encourages a transition towards a more covariant approach to quantum field theory, whereby degrees of freedom are understood as existing over regions of spacetime, as is natural in a theory of quantum gravity. This can also be seen in the definition of the SJ state, which we discuss in Appendix \ref{ap: SJ} and which we use in our example in Section \ref{sec: ee example}. Similar to the entropy definition, the SJ state is covariantly and spectrally defined through modes characterised by the entire spacetime. This is advantageous over traditional methods for defining a state which use auxiliary structures, such as Killing fields and Hamiltonians, which are not covariant or do not necessarily exist in arbitrary curved or discrete spacetimes.

Another unique aspect of our work, is that we have treated both bosonic and fermionic theories in a closely parallel manner. This has helped bring out some interesting similarities and differences between them. For example, in both theories it is the spacetime two-point (Wightman) correlation function and causal propagator ($G_R-G_A$) that determine the entropy.\footnote{As we are dealing with quasifree theories, which are fully specified by their spacetime two-point correlation function, it is intuitive and expected that the entropy must also be fully captured by this object. Our formulae illustrate how this expectation is realised. (The causal propagator itself can also be written in terms of the two-point correlation function).} The causal propagator is directly related to the spacetime field commutator in the bosonic theory, and the anticommutator in the fermionic theory. Hence we see the appearance and role of the relevant spin-statistics of each theory in quantifying the entropy. In the fermionic case, the generalised eigenvalues that enter the entropy sum are all positive and equal to the familiar eigenvalues of the density matrix; in the bosonic theory the generalised eigenvalues are less straightforwardly related to the eigenvalues of the density matrix (though we derive their precise relation) and they can be both positive and negative. The parallel treatment of the SJ state in both theories also sharpens their comparison. For example, both states are spectrally, covariantly, and uniquely defined. However, central to the bosonic construction is the manifestly spacetime $L^2$ (rather than the hypersurface-based Klein-Gordon) inner product, whereas the Dirac (as well as the analogue of the $L^2$) inner product plays an essential role in the fermionic construction.

Our ability to perform entropy calculations over spacetime regions opens up many avenues for investigation into the nature and identity of the degrees of freedom of quantum field theories and quantum gravity. Some areas of further exploration include the extension of this formalism to other quasifree field theories, interacting theories (beyond the perturbative first order, which has been studied), and relative entropy. Most importantly, more applications of this formalism are needed to see its strengths in use and to discover the subtleties and lessons that may arise from such a spacetime treatment of field degrees of freedom. This will also continue the progress towards finding out whether the entanglement of spacetime field degrees of freedom is a viable source of black hole entropy, and in fact, all
horizon entropy. In future work we aim to carry out such investigations.

\acknowledgments
This work was conducted with the support of Research Ireland under grant number 22/PATH-S/10704. 

\appendix

\section{Thermal Entropy Calculations in the Continuum} \label{examplesec}

In this appendix we show two examples that serve as a proof of concept that the formulae presented in Sections \ref{bosonsec} and \ref{fermsec} reproduce known results. These examples are: the thermal entropy of a real scalar and Dirac fermion field in $(3+1)$-dimensional Minkowski spacetime.

\subsection{Thermal Entropy of a Real Scalar Field in $\textbf{R}^{(3,1)}$}

We will first calculate the thermal entropy of a real scalar field in $(3+1)$-dimensional Minkowski spacetime using the expressions  in Section \ref{bosonsec}. Our setup, and procedure, will in fact be similar to the one of that section. A related calculation in $0+1$ dimensions can be found in Appendix B of \cite{SSY2014}.

We will work with the expansion of the field in terms of the standard modes that have positive and negative frequency with respect to the static time coordinate of Minkowski spacetime
\begin{equation}
    \phi(x) = \int \frac{d^{3}p}{(2 \pi)^{3}} \frac{1}{\sqrt{2\omega_{\vec{p}}}} \left(a^{\dagger}_{\vec{p}}e^{-ip\cdot x} + a_{\vec{p}}e^{ip \cdot x}\right),
\end{equation}
where the $4$-momenta are on-shell, i.e. $\omega_{\vec{p}}= \sqrt{\vec{p}\cdot\vec{p}+m^{2}}$. This basis is orthogonal with respect to both the Klein-Gordon inner product and the $L^{2}$ spacetime inner product (see Appendix \ref{sec: BIP} where this is discussed in more detail). The thermal density matrix is 
\begin{equation}
    \rho = \frac{e^{-\beta \mathcal H}}{\text{tr}\left(e^{-\beta \mathcal H}\right)},\qquad \mathcal H = \int \frac{d^{3}p}{(2\pi)^{3}}\ \omega_{\vec{p}}\, a^{\dagger}_{\vec{p}}\, a_{\vec{p}}\,, 
\end{equation}
where $\beta=\frac{1}{T}$. We can thus obtain the thermal two-point function and commutator, 
\begin{gather}
   \scalebox{0.99}{$\displaystyle\langle\Psi|\phi(x) \phi(x')|\Psi\rangle_{\beta} = \frac{1}{({2 \pi})^{3}}\int d^{3}p \frac{1}{2\omega_{\vec{p}}} \left(\frac{1}{1-e^{-\beta \omega_{\vec{p}}}}e^{i p\cdot (x-x')} + \frac{e^{-\beta \omega_{\vec{p}}}}{1 - e^{-\beta \omega_{\vec{p}}}}e^{-i p\cdot (x-x')} \right),$}
   \\
    [\phi(x), \phi(x')] = \frac{1}{({2 \pi})^{3}}\int d^{3}p \frac{1}{2\omega_{\vec{p}}} \left(e^{i p\cdot (x-x')} - e^{-i p\cdot (x-x')} \right). 
\end{gather}
From the expressions above, we can directly read off the $A_{\vec{p}}$ and $B_{\vec{p}}$ coefficients needed to calculate the entropy. These are
\begin{equation}
    A_{\vec{p}} = \frac{1}{1 - e^{-\beta \omega_{\vec{p}}}}, \qquad B_{\vec{p}}= \frac{e^{-\beta \omega_{\vec{p}}}}{1 - e^{-\beta \omega_{\vec{p}}}}.\label{eq: A B thermal scalar}
\end{equation} 
We may now insert \eqref{eq: A B thermal scalar} into \eqref{Deltaexpr} and \eqref{eq: full entropy} (which is equivalent to our formula \eqref{Ssum}), noting that we must perform an integral instead of a sum since the index $\vec{p}$ is continuous. To understand the relation between the sum and integral, we can use a box regularisation that is sent to infinity, 
\begin{equation}
    \sum_{\vec{p}} = \left(\frac{ L} {2\pi}\right)^3 \sum_{\vec{p}}(\Delta p)^3 \xrightarrow{L\to \infty}  L^3\int \frac{d^{3}p}{(2 \pi)^{3}} = V\int \frac{d^{3}p}{(2 \pi)^{3}},
\end{equation}
where $L$ is the side length of the box and $\Delta p=\frac{2 \pi}{L}$ is the difference between neighbouring wavenumbers. We can thus calculate an entropy density 
\begin{equation}
\begin{split}
    s = \frac{S}{V} &= \int \frac{d^{3}p}{(2\pi)^{3}} \left(\frac{1}{1 - e^{-\beta \omega_{\vec{p}}}} \ln\left(\frac{1}{1 - e^{-\beta \omega_{\vec{p}}}}\right) -\frac{e^{-\beta \omega_{\vec{p}}}}{1 - e^{-\beta \omega_{\vec{p}}}} \ln \left(\frac{e^{-\beta \omega_{\vec{p}}}}{1 - e^{-\beta \omega_{\vec{p}}}}\right)\right) \label{thermalexpr}\\
    &= \int \frac{d^{3}p}{(2\pi)^{3}} \left(\frac{\beta \omega_{\vec{p}}}{e^{\beta \omega_{\vec{p}}}-1} -\ln{\left(1 - e^{-\beta \omega_{\vec{p}}} \right)}\right),
\end{split}
\end{equation}
showing that each mode contributes the entropy of a thermal harmonic oscillator. The same result can be obtained (see e.g. Section 2.1 of  \cite{basics.pdf}) using the standard  relation 
\begin{equation}
    S = \frac{\partial (T \ln Z)}{\partial T},
\end{equation}
where $Z$ is the thermal partition function
\begin{equation}
    Z =  \prod_{{\vec{p}}}\exp{\left(-\frac{\omega_{\vec{p}}}{2 T} - \ln\left(1-e^{-\left(\frac{\omega_{\vec{p}}}{T}\right)}\right) \right)}.
\end{equation}
Taking a massless field, such that $\omega_{\vec{p}}=|\vec{p}|=:p$, \eqref{thermalexpr} becomes
\begin{equation}
\begin{split}
    s &= \frac{1}{(2\pi)^{3}}\int^{\infty}_{0} dp (4 \pi p^{2}) \left(\frac{\beta p}{e^{\beta p}-1} -\ln{\left(1 - e^{-\beta p} \right)}\right), \\
    &= \frac{2 \pi^{2}}{45 \beta^{3}},
\end{split}
\end{equation}
the expected answer \cite{basics.pdf}. For a massive field, while the entropy density is finite, a closed form expression is not known (to our knowledge). However, one can in this case either numerically evaluate the entropy density, or analytically write out the high and low temperature expansions  \cite{basics.pdf}.

\subsection{Thermal Entropy of a Dirac Fermion Field in $\textbf{R}^{(3,1)}$} \label{sec:thermalferm}
Now we will repeat the example of the previous subsection, but for a fermion field. We treat the  Dirac field in a thermal state, because it is the most common example. We will decompose the Dirac field into two Majorana fields as we mentioned is possible earlier.\footnote{Strictly speaking one does not need to do so in this case. Using a Hamiltonian with charge conjugation symmetry means the same treatment for Majorana fermions also applies to other fermion species, provided one factors in the differing number of internal degrees of freedom. This is because the symmetry ensures the diagonal Hamiltonian has the same scalar coefficient for particles and antiparticles.}

Once again, we begin with the standard mode expansion of the field in terms of positive and negative frequency modes with respect to the static time coordinate of Minkowski spacetime
\begin{equation} \label{eq: fermion psi}
    \psi(x) = \int \frac{d^{3}p}{(2\pi)^{3}} \frac{1}{\sqrt{2 \omega_{\vec{p}}}} \left(a^s_{\vec{p}} u^{s}(\vec{p}) e^{ip\cdot x} + b^{s\dagger}_{p} v^{s}(\vec{p}) e^{-ip\cdot x} \right),
\end{equation}
where summation over $s$ is implied. Here $u^{s}(\vec{p})$ and $v^{s}(\vec{p})$ are Dirac basis spinors. The basis in \eqref{eq: fermion psi} is orthogonal with respect to the conventional Dirac inner product.
The thermal density matrix is
\begin{equation}
    \rho = \frac{e^{-\beta \mathcal H}}{\text{tr}\left(e^{-\beta \mathcal H}\right)},\qquad \mathcal H = \int \frac{d^{3}p}{(2\pi)^{3}}\ \omega_{\vec{p}}\, (a^{s\dagger}_{\vec{p}}\, a^s_{\vec{p}}\, + b^{s\dagger}_{\vec{p}}\, b^s_{\vec{p}}\,).
\end{equation}

Let us now proceed to decompose the Dirac fermion into two Majorana fermions.\footnote{Here we will need to make the identification $C\bar{u}^{T} = v$, $C\bar{v}^{T} = u$. This is true up to a phase, which we choose to be trivial. } We define the appropriate combinations
\begin{align}
    \chi_{1}(x) := \frac{1}{\sqrt{2}}\left( \psi(x)+\psi^* (x)\right), \\
    \chi_{2}(x) := \frac{i}{\sqrt{2}}\left( \psi(x)-\psi^* (x)\right),
\end{align}
where again $^*$ denotes the field under charge conjugation. We then have mode expansions for our Majorana fermions 
\begin{equation}
    \chi_{i}(x) = \int \frac{d^3 p}{(2\pi)^3} \frac{1}{\sqrt{2 \omega_{\vec{p}}}} \left( c^{s}_{i,\vec{p}} u^{s}(\vec{p}) e^{i p \cdot x} + c^{\dagger}_{i,\vec{p}} (u^{s})^{*}(\vec{p}) e^{-i p \cdot x} \right),
\end{equation}
wherein we have defined
\begin{equation}
    c_{1,\vec{p}}^{s} := \frac{1}{\sqrt{2}}(a_{\vec{p}}^{s}+b_{\vec{p}}^{s}), \quad c_{2,\vec{p}}^{s} := \frac{i}{\sqrt{2}}(a_{\vec{p}}^{s}-b_{\vec{p}}^{s}),
\end{equation}
which inherit the relations
\begin{equation}
    \{ c^{r}_{i,\vec{p}}, c^{s\dagger}_{j,\vec{p}\,'} \} = (2\pi)^3 \delta^{3}(\vec{p} - \vec{p}\,') \delta_{ij}\delta^{rs}.
\end{equation}
Using our definitions for the $c_{i,\vec{p}}^{s}$, we find that the Hamiltonian decomposes as
\begin{equation}
    \mathcal{H} = \mathcal{H}_1 + \mathcal{H}_2, \quad \mathcal H_{i} = \int \frac{d^{3}p}{(2\pi)^{3}}\, \omega_{\vec{p}}\, c^{s\dagger}_{i,\vec{p}}\, c^s_{i,\vec{p}}.
\end{equation}
As our Hamiltonians act in different subspaces of our Fock space, and commute, due to containing only the number operator, the density operator factorises, as 
\begin{equation}
    \rho = \frac{e^{-\beta \mathcal{H}}}{\mathrm{Tr}\left(e^{-\beta \mathcal{H}}\right)} = \rho_1 \otimes \rho_2, \quad \rho_i = \frac{e^{-\beta \mathcal{H}_i}}{\mathrm{Tr}\left(e^{-\beta \mathcal{H}_i}\right)}.
\end{equation}
We thus have the exact same structure for each Majorana fermion. Taking one of them, the thermal two-point function and anticommutator are 
\begin{gather}
\begin{split}
    \langle \Psi | \chi_{i}(x) \overline{\chi}_{i}(x') |\Psi \rangle_{\beta} = \frac{1}{(2\pi)^{3}}\int d^{3}p \frac{1}{2 \omega_{\vec{p}}} &\Bigl( \frac{1}{1+e^{-\beta \omega_{\vec{p}}}}u^{s}(\vec{p})\bar{u}^{s}(\vec{p}) e^{ip\cdot (x-x')} \\
    &+ \frac{e^{-\beta \omega_{\vec{p}}}}{1+e^{-\beta \omega_{\vec{p}}}} u^{s*}(\vec{p})\bar{u}^{s*}(\vec{p}) e^{-ip\cdot (x-x')} \Bigl), \end{split}\\
    \scalebox{0.99}{$\displaystyle\{\chi_{i}(x), \overline{\chi}_{i}(x')\} = \frac{1}{(2\pi)^{3}}\int d^{3}p \frac{1}{2 \omega_{\vec{p}}} \left(u^{s}(\vec{p})\bar{u}^{s}(\vec{p}) e^{ip\cdot (x-x')} + u^{s*}(\vec{p})\bar{u}^{s*}(\vec{p})e^{-ip\cdot (x-x')} \right).$}
\end{gather}

From the expressions above, we can again immediately read off the probabilities \eqref{eq: pj0 fermion} and \eqref{eq: pj1 fermion} needed to compute the entropy. These are 
\begin{equation}\label{eq: fermion probabilities}
    p_{\vec{p},0} = \frac{1}{1+e^{-\beta \omega_{\vec{p}}}},\qquad  p_{\vec{p},1} = \frac{e^{-\beta \omega_{\vec{p}}}}{1+e^{-\beta \omega_{\vec{p}}}}.
\end{equation}
We can then insert \eqref{eq: fermion probabilities} into the sum \eqref{eq: fermion mode entropy} (equivalent to our formula \eqref{Ssumf}) and integrate over the components of $\vec{p}$, taking into account the twofold degeneracy of the spectrum introduced by the $u^{s}(\vec{p})\bar{u}^{s}(\vec{p})$ and $u^{s*}(\vec{p})\bar{u}^{s*}(\vec{p})$ operators. As in the previous subsection, we compute the entropy density for our Dirac fermion as
\begin{equation}
\begin{split}\label{eq: fermion entropy density}
    s &= -2\times2\int^{\infty}_{0} \frac{d^{3}p}{(2\pi)^{3}} \left(\frac{1}{1+e^{-\beta \omega_{\vec{p}}}} \ln \left(\frac{1}{1+e^{-\beta \omega_{\vec{p}}}}\right) + \frac{e^{-\beta \omega_{\vec{p}}}}{1+e^{-\beta \omega_{\vec{p}}}} \ln\left(\frac{e^{-\beta \omega_{\vec{p}}}}{1+e^{-\beta \omega_{\vec{p}}}}\right)\right) \\
    &= 4\int^{\infty}_{0} \frac{d^{3}p}{(2\pi)^{3}} \left( \frac{\beta \omega_{\vec{p}}}{e^{\beta \omega_{\vec{p}}}+1} + \ln\left( 1+e^{-\beta \omega_{\vec{p}}}\right)\right),
\end{split}
\end{equation}
where the first factor of $2$ is from the two Majorana fermions, and the second is from the degeneracy of the $u^{s}(\vec{p})\bar{u}^{s}(\vec{p})$ and $u^{s*}(\vec{p})\bar{u}^{s*}(\vec{p})$ operators. Once again we see that each mode contributes an individual thermal entropy, this time that of a fermionic harmonic oscillator. For a massless Dirac field, \eqref{eq: fermion entropy density} becomes
\begin{equation}
\begin{split}
    s &= \frac{4}{(2\pi)^{3}}\int^{\infty}_{0} dp (4 \pi p^{2}) \left( \frac{\beta p}{e^{\beta p}+1} + \ln\left( 1+e^{-\beta p}\right)\right), \\
    &= \frac{7 \pi^{2}}{45 \beta^{3}},
\end{split}
\end{equation}
which is again in agreement with the literature \cite{basics.pdf}. As in the scalar case, the massive fermion field also has a finite thermal entropy density, but one for which a closed form is not known.

\section{The Sorkin-Johnston State}\label{ap: SJ}

The Sorkin-Johnston (SJ) state \cite{Johnston2009, Sorkin2011, FGFTQF} is a spectral, covariant and distinguished vacuum state that takes into consideration the full spacetime. It is defined in a hypersurface-free way, as is fitting for a frame independent formulation. This makes it a natural state to use for studying the spacetime entropy we have been discussing in this work, which similarly has a spectral and covariant character. The spacetime nature of the SJ state also makes it particularly suited to causal sets, wherein it assumes the status of canonical vacuum. In static spacetimes of infinite temporal extent, the bosonic SJ state matches the  familiar vacuum state selected by the hypersurface-orthogonal timelike Killing field, a point we will return to in Section \ref{sec: BIP}. In the fermionic case, the conventional static vacuum must satisfy  an additional constraint (which we discuss in Section \ref{sec: FIP}) to be the same as the SJ state. In more general spacetimes, e.g. ones without any symmetries or compact ones,  a distinguished vacuum reflecting the properties of the spacetime is still selected by the SJ prescription. For some examples of the SJ state, see \cite{Aslanbeigi:2013fga, Buck:2016ehk, Surya:2018byh, Mathur:2019yvl}.
\subsection{The SJ Prescription}
We review the definition of the SJ state for  bosonic theories  in Section \ref{sec: sj boson}, and the analogous definition for  fermionic theories in Section \ref{sec: sj ferm}. It is worth briefly pointing out that to guarantee the existence of these constructions, mathematical considerations require that we work with compact spacetimes, or alternatively Hilbert spaces of functions with compact spacetime support, to ensure that  the spacetime inner product is bounded.
\subsubsection{The SJ State for Bosons}\label{sec: sj boson}
To construct the SJ state, we begin with the retarded and advanced Green functions, $G_{R,A}$, which were defined in \eqref{eq: gr ga}. From these, we construct the Pauli-Jordan function (or causal propagator) via the Peierls relation,

\begin{equation} \label{eq: pj function}
     \Delta(x,x') = G_{R}(x,x')-G_{A}(x,x').
\end{equation}
Multiplying \eqref{eq: pj function} by a factor of $i$ we obtain the spacetime commutator ($i\Delta$) that we encountered in \eqref{eq: iDelta commutator}. With the Pauli-Jordan function at hand, the SJ state is then defined via the Wightman function that is constructed from the positive eigenvalues and corresponding eigenfunctions of $i\Delta$, with respect to the $L^2$ inner product over the spacetime.

We can explain this more directly as follows. Using  the $L^2$ inner product on the spacetime manifold $\mathcal M$, 
\begin{equation}
    \llangle h , f\rrangle_{L^2} := \int_{\mathcal M} h^{*}(x)f(x) dV,
\end{equation}
we promote the Pauli-Jordan function to an operator\footnote{We will omit the hats from our operators in the appendices, for  notational simplicity.} (as we did in \eqref{eq: Delta as operator}), and take its positive spectral part. We can always do this in bounded spacetime regions, as there the resultant operator is self-adjoint \cite{FV2012}.\footnote{An unbounded spacetime can often be treated as a limiting case in which a bounded spacetime has been extended to infinity (as done in \cite{AAS2012}), or by otherwise assigning to $i\Delta$ a (unique) self-adjoint extension (when possible).} The integral kernel of this positive operator is then the SJ Wightman two-point function for the SJ vacuum state. This is often denoted by 
\begin{equation}
    W_{SJ} := \text{pos}(i\Delta),
\end{equation}
which in a causal set, is implemented as
\begin{equation}
    i\Delta = \sum_{i} \tilde\lambda_i v_{i} v^{\dagger}_{i} \,\implies\, W_{SJ} = \sum_{i} \Theta(\tilde\lambda_i) \tilde\lambda_i v_{i} v^{\dagger}_{i},
\end{equation}
where $\{v_i\}$ and $\{\tilde\lambda_i\}$ are the $L^2$-normalised eigenfunctions and eigenvalues of $i\Delta$, respectively. 

As an alternative to the above construction, it can be shown that the necessary conditions of purity, positive semidefiniteness, and hermiticity uniquely specify the vacuum state up to a choice of inner product. The $L^2$ spacetime inner product then serves as the natural choice of inner product, due to its generality, covariance, and respect for spacetime symmetries. We refer the reader to \cite{JJ2024} for an axiomatic derivation of the SJ state along these lines, and we present a sketch of the argument below.

Let us begin with our entropy formula, \eqref{Ssum}, and impose the condition of purity: $S(W)=0$. We use the $L^2$ spacetime inner product to promote our bilinear objects to operators, for the reasons mentioned above. That the entropy vanishes means that the solutions $\lambda$ to the generalised eigenvalue problem \eqref{Sev} must be $0$ or $1$. This makes $-i\Delta^{-1} W$ a projection operator\footnote{where it is understood that we are working in the subspace where $\Delta$ is invertible.}, and as such 
\begin{equation} \label{eq: projection}
    W\Delta^{-1}W = iW.
\end{equation}
Expanding \eqref{eq: projection} by writing $W$ in terms of its real and imaginary parts, using the fact that $\text{Im}(W) = \frac{\Delta}2$, and taking $W$ and $i\Delta$ to be simultaneously diagonalisable and therefore commute, it can be shown that
\begin{equation}
    \text{Re}(W)=\frac{1}{2} \sqrt{-\Delta^2},
\end{equation}
and so
\begin{equation} \label{eq: w positive part of iDelta}
    W=\frac{1}{2}\left(\sqrt{-\Delta^2}+i\Delta\right)=\text{pos}(i\Delta).
\end{equation}

As $i\Delta$, and hence $W$, can be defined directly from the retarded and advanced Green functions, global hyperbolicity is a sufficient condition for the existence and uniqueness of the SJ state. The same cannot be said for vacua defined via Killing fields.

Finally, we remark that the SJ state does not always have the Hadamard property\footnote{The Hadamard property imposes that a vacuum state's singularity structure should match that of Minkowski spacetime. This can be stated in terms of the two-point function having the form 
\begin{equation}
    W(x, x')  = \lim_{\epsilon\to0^{+}} \alpha_{d}\left(\frac{U(x,x')}{(\sigma(x,x')+i\epsilon(t-t'))^{\frac{d-2}{2}}} + V(x,x') \ln\left(\frac{\sigma(x,x')+i\epsilon(t-t')}{l^2}\right) + F(x,x')\right).
\end{equation}
Here $\alpha_{d}$ is a dimension dependent constant, $\sigma(x,x')$ is the Synge world function, $l$ is some length scale, and $U(x,x')$, $V(x,x')$, and $F(x,x')$ are all regular symmetric bilinear functions with expansions in powers of the Synge world function; see e.g. \cite{WaldQFTCS}.} \cite{FV2012, Zhu:2022kcf} in continuum settings. In several known examples where the non-Hadamard behaviour is observed \cite{FV2012, FL15}, it is an artifact of the bounded temporal extent of the spacetime. In such cases, simple modifications can restore the Hadamard condition, if one is concerned about this. This is done via a softening of the (spacelike or null) boundary \cite{FV2013, BF13, FGFTQF}, at the cost of uniqueness, given that the choice of softening function is not unique. The Hadamard condition is not meaningful in discrete settings, and no such condition is necessary for the purposes of (e.g. point-splitting) regularisation, for which it is commonly used \cite{Fulling:1978ht}, as the discreteness automatically regulates field quantities.

\subsubsection{The SJ State for Fermions}\label{sec: sj ferm}
Constructing the SJ state for fermions is conceptually the same, yet a mathematical subtlety makes its treatment more tortuous. The subtlety is that while the ${L^{2}}$ spacetime inner product of the bosonic theory is positive definite, the analogous fermionic spacetime $L^{2}$ inner product is \emph{indefinite}. The usual known spectral theorem applies to self-adjoint (or more generally, normal) operators on a space with a positive definite inner product. There is no known analogous spectral theorem for indefinite inner product spaces, except for a subset of operators, and when the indefinite inner product is related to a positive definite inner product\footnote{via a fundamental symmetry operator, $J$, such that $\llangle\cdot, J \cdot\rrangle>0$. Such a space is known as a Krein space.} \cite{Bognar_1974}. One such class of operators are ``definitisable" (or ``positisable") operators \cite{Langer82, Jonas81}. Another such class of operators are those which are similar to a self-adjoint operator in a Hilbert space\footnote{or equivalently, fundamentally reducible operators.} \cite{McEnnis1982}. Alternatively, one may find that in a relevant subspace, one's operator of interest becomes a self-adjoint operator in a Hilbert space, and the usual spectral theorem applies. The relation between the $L^2$ inner product and the positive definite Dirac inner product makes this the case for $i\Delta_F$, and we will use this to spectrally decompose the anticommutator.\footnote{In the bosonic theory it was the spacetime commutator that was spectrally decomposed. Here we must use the spacetime anticommutator.} Although our presentation is different, the SJ state we construct is the same as that constructed in \cite{FL15}, based on the fermionic projector approach of  \cite{Finster2013}. The SJ state differs from the fermionic projector state, as the SJ state is the projection into the positive spectral subspace of $i\Delta_F$, whereas the fermionic projector state is the negative of the projection into the negative spectral subspace of $i\Delta_F$.

Let us lay out again the two relevant inner products, which we already encountered in Section \ref{fermsec}. The Dirac (hypersurface) inner product is 
\begin{equation}\label{eq: dirac hypersurface inner product}
    \llangle \varpsi, \tilde{\varpsi} \rrangle_D = i\int_{\Sigma} \bar{\varpsi}(x)\, \slashed{n}\, \tilde{\varpsi}(x)\, dS,
\end{equation}
where the integration is over the Cauchy surface $\Sigma$ and $n^\mu$ is orthogonal to it. The fermionic equivalent of the spacetime  ${L^{2}}$ inner product is 
\begin{equation}\label{eq: fermion L2 inner product}
    \llangle f, h \rrangle_{L^{2}} = \int_{\mathcal M} \bar{f}(x)\, h(x)\, dV.
\end{equation}
As we saw in \eqref{eq: relation between D and L2}, these two inner products are related by\footnote{This relation is shown in Proposition 2.2 of \cite{Dimock1982DiracQF}, and is a special case of a corollary of the Riesz lemma (see e.g.  page $44$ of \cite{ReedSimon}).}
\begin{equation}\label{eq: relation between D and L2 appendix}
    \llangle \varpsi, \tilde{\varpsi} \rrangle_D  = -i\llangle \varpsi, h \rrangle_{L^{2}},
\end{equation}
where $\varpsi$ and $\tilde\varpsi$ are solutions to the Dirac equation, and $h$ is a test spinor field (element of $C^{\infty}_{0}(D\mathcal M)$) such that $\tilde{\varpsi} = \hat\Delta_{F} h$. We can apply  
\eqref{eq: relation between D and L2 appendix} to solutions, 

 \begin{equation}\label{eq: relation between L2 and D appendix}
    \llangle \varpsi, \tilde{\varpsi} \rrangle_{L^{2}}  = \llangle \varpsi, i\hat\Delta_F  \tilde{\varpsi} \rrangle_{D},
\end{equation}
where $\tilde{\varpsi}$ now has compact spacetime support. Therefore if we work with the space of solutions to the Dirac equation, which is equivalent to the subspace of $L^2$ that is in the image of $\Delta_F$ (which is the space of interest, after all), \eqref{eq: relation between L2 and D appendix} holds. With this relation to a positive definite inner product,  we can proceed\footnote{This relation between the inner products allows us to express the matrix elements $(i{\Delta_F})_{ij}$ as $\llangle \varpsi_i, i\Delta_F \varpsi_j\rrangle_D = \llangle \varpsi_i, \varpsi_j\rrangle_{L^2}$. Here $\{\varpsi_i\}$ is the basis of solutions of the Hilbert space with the Dirac inner product, which we use to represent $i\Delta_F$, and diagonalise it.} to find the spectral decomposition of  $i\Delta_F$ and take its positive part to define the SJ state, 
\begin{equation}\label{eq: fermionic WSJ}
    W_{SJ} := \text{pos}(i\Delta_F).
\end{equation}

We can see that $W_{SJ}$ is positive semidefinite by construction, as is required for a Wightman function. $W_{SJ}$ also leads to a pure state, as its nonzero spectrum will be a subset of the spectrum of $i\Delta_{F}$, and so, upon insertion into our generalised eigenvalue formula \eqref{eq: gen eig ferm}, must return eigenvalues of $0$ or $1$, resulting in $S=0$.

Although the fermionic SJ prescription requires a slightly different treatment than its bosonic counterpart, it is in spirit the same. In both cases, we obtain spectrally (from each theory's relevant (anti)commutation relation) a natural state that respects spacetime symmetries and covariance. In the bosonic theory, we were able to use the manifestly frame independent and positive definite $L^2$ inner product to do this. In the fermionic theory we have to use the positive definite hypersurface sesquilinear form induced by the Dirac current together with the analogue of the $L^2$ inner product, since the latter is indefinite. While it may seem that using the Dirac inner product introduces some frame dependence, this is not in fact the case. The Dirac inner product, while defined on a Cauchy hypersurface and therefore less evidently covariant than the spacetime $L^2$ inner product, is nevertheless hypersurface independent \cite{Dimock1982DiracQF}.\footnote{This can also be seen from the spacetime representation of this inner product, as in \eqref{eq: relation between D and L2 appendix}.} It is only the use of this inner product in conjunction with conventional nonspectral methods for picking a positive subspace, such as putting a Hamiltonian in its ground state at a moment in time, that  introduces frame dependence.  Here, the SJ prescription takes care of this in a frame independent manner by using the spectrum of  $i\Delta_F$ together with the covariant $L^2$ inner product.

\subsection{Static Spacetimes} \label{sec: static} 
In this subsection, we show that in static spacetimes, the bosonic SJ state agrees with the conventional vacuum state that is distinguished by the Killing field, while the fermionic SJ state does so if the conventional static vacuum satisfies an additional constraint. A similar discussion as subsection \ref{sec: BIP}  appears in \cite{AAS2012} and \cite{DiamondGS}. 
\subsubsection{Bosonic Case}
\label{sec: BIP}
The defining condition of the bosonic SJ state,  \eqref{eq: w positive part of iDelta}, is equivalent to $W$ and $W^*$ having orthogonal support \cite{FGFTQF}, i.e.
\begin{equation}\label{eq: orthogonal support}
    WW^T=0.
\end{equation}
This condition is satisfied by a Wightman function $W(x,x')=\sum\limits_{i} f_i(x)f_i^*(x')$ if $\llangle f_{i},f_{j}^{*}\rrangle_{L^2}=0$. Hence if we have a Wightman function obtained through other means, we can find out whether or not it is the SJ state by checking this orthogonality condition. Below, we carry out this exercise for the conventional Wightman functions of static spacetimes.

If a spacetime is static, it has an everywhere timelike and hypersurface-orthogonal Killing vector, $K^{\mu}$. We can choose coordinates such that the metric takes the form
\begin{equation}
    ds^{2} = -\alpha(\vec{x})dt^{2} + h_{i j}(\vec{x})dx^{i}dx^{j},
\end{equation}
where $h_{i j}$ specifies a Riemannian metric. In these coordinates, the Klein-Gordon equation is
\begin{equation}
    \left(\partial_{t}^{2} - H(\vec{x})\right)\varphi(x)=0; \quad H(\vec{x}):=\alpha(\vec{x})\left[\frac{1}{\sqrt{-g(\vec{x})}}\partial_{i}\left(\sqrt{-g(\vec{x})}\,h^{ij}(\vec{x})\partial_{j}\right)-m^{2}\right].
\end{equation}
The appropriate inner product to consider on a Cauchy surface is
\begin{equation}
    \llangle f,\tilde{f} \rrangle_{\Sigma} := \int_{\Sigma} \frac{1}{\sqrt{\alpha(\vec{x})}} f^{*}(\vec{x})\tilde{f}(\vec{x}) \sqrt{h}\, d^{d-1}x,
\end{equation}
and in this inner product, $H$ is negative semidefinite and formally self-adjoint. This ensures a spectral theorem, and so we can write an eigenvalue equation of the form 
\begin{equation}
    (H g_{p})(\vec{x}) = - \omega_p^{2}\, g_{p}(\vec{x}).
\end{equation}
We can then find a set of solutions $\{f_p,f^*_p\}$ to the Klein-Gordon equation, where 
\begin{equation}
    f_{p}(x) = \frac{1}{\sqrt{2\, \omega_p}}g_{p}(\vec{x})e^{- i\omega_p t}, \label{KGsolns}
\end{equation}
with $\omega_p>0$. These positive frequency solutions will form a basis for the one-particle Hilbert space. We would like to now find a condition such that these solutions are $\delta$-function normalised in the Klein-Gordon inner product. We can define a Cauchy surface $\Sigma$ with $n^{\mu} := \frac{K^{\mu}}{|K|}= \frac{\delta^{\mu}_{t}}{\sqrt{\alpha(\vec{x})}}$ as its unit normal everywhere, and hence  the Klein-Gordon inner product on this Cauchy surface becomes
\begin{equation}\label{eq: KG orthogonality of modes}
\begin{split}
    \llangle f_{p} , f_{p'} \rrangle_{KG} &= i\int_{\Sigma}\left \{f^{*}_{p}(x) n^{\mu} \nabla_{\mu} f_{p'}(x) - f_{p'}(x) n^{\mu} \nabla_{\mu} f^{*}_{p}(x) \right\}\sqrt{h}\, d^{d-1}x \\
    &= \frac{\omega_p+\omega_{p'}}{\sqrt{4\,\omega_p\,\omega_{p'}}} e^{ -i \left(\omega_{p'}-\omega_p\right)t} \int_{\Sigma} \frac{1}{\sqrt{\alpha(\vec{x})}} g_{p}^{*}(\vec{x})g_{p'}(\vec{x}) \sqrt{h}\, d^{d-1}x.
\end{split}
\end{equation}
We thus see that provided that we choose our spatial components to be appropriately normalised in the spatial integral, i.e. that 
\begin{equation}
    \llangle g_{p},g_{p'}\rrangle_{\Sigma} = \delta^{d-1}(\vec{p}-\vec{p}\,'),
\end{equation}
our chosen set of solutions forms a basis of the solution space of the Klein-Gordon equation, orthonormal with respect to the Klein-Gordon inner product. This choice of positive modes  defines a vacuum state.

Let us now consider the $L^{2}$ inner product, and check whether our choice of solutions to the Klein-Gordon equation, (\ref{KGsolns}), is orthogonal with respect to this inner product as well. We have 
\begin{equation}\label{eq: kg sols in L2}
\begin{split}
    \llangle f_{p}, f_{p'}\rrangle_{L^{2}} =& \int_{\mathcal M}  f^{*}_{p}(x)f_{p'}(x) \sqrt{-g}\, d^{d}x \\ 
    =& \frac{2 \pi}{\sqrt{4\,\omega_p\,\omega_{p'}}} \int \frac{e^{ -i (\omega_{p'}-\omega_p) t}} {2 \pi}dt \int \sqrt{\alpha(\vec{x})} g^{*}_{p}(\vec{x}) g_{p'}(\vec{x}) \sqrt{h}\, d^{d-1}x \\
    =& \frac{2 \pi}{\sqrt{4\,\omega_p\, \omega_{p'}}} \delta(\omega_p-\omega_{p'}) \int \sqrt{\alpha(\vec{x})} g^{*}_{p}(\vec{x}) g_{p'}(\vec{x}) \sqrt{h}\, d^{d-1}x,
\end{split}
\end{equation}
where we have taken the limits of the $t$ integral to $\pm\infty$. We see that the integral we are left with in \eqref{eq: kg sols in L2} does not agree with our previous spatial integral in \eqref{eq: KG orthogonality of modes}, unless $\alpha=\text{const}$. Therefore, functions that are Klein-Gordon orthogonal will not always be $L^{2}$ orthogonal as well, since $\omega_p$ and $\omega_{p'}$ can be equal for $p\neq p'$. Crucially, however, positive and negative frequency modes remain orthogonal under the $L^{2}$ inner product: 

\begin{equation}
    \llangle f^*_{p}, f_{p'}\rrangle_{L^{2}} 
    = \frac{2 \pi}{\sqrt{4\,\omega_p\, \omega_{p'}}} \delta(\omega_p+\omega_{p'}) \int \sqrt{\alpha(\vec{x})} g_{p}(\vec{x}) g_{p'}(\vec{x}) \sqrt{h}\, d^{d-1}x=0,
\end{equation}
 since $\omega_p,\omega_{p'}>0$.  This means that the SJ state (defined using the $L^2$ inner product) in static spacetimes will be the same one that is picked out by $K^\mu$ and the Klein-Gordon inner product.
 
\subsubsection{Fermionic Case} \label{sec: FIP}
In a static spacetime, we can express the Dirac equation as \cite{Jin:2000ra}
\begin{equation}
  \left(i \partial_{t} - H(\vec{x}) \right) \varpsi(x)=0; \quad H(\vec{x}) :=\alpha(\vec{x}) i \gamma^{0}\gamma^{i}\nabla_{i} - i\alpha(\vec{x}) m \gamma^{0} - i\Gamma_{0}(\vec{x}),
\end{equation}
where $\nabla_{\mu}$ is the covariant derivative. Proceeding as in the bosonic case, we can then write an eigenvalue equation 
\begin{equation}
    (H g^\pm_{p})(\vec{x}) =\pm \omega_p\, g^\pm_{p}(\vec{x}),
\end{equation}
where $\omega_p>0$. This allows us to define a basis of the solution space,
in terms of positive and negative frequency modes
\begin{align}\label{Diracsolns}
    f^{+}_{p}(x) &= g^+_{p}(\vec{x})e^{ -i\omega_p t}=:f_{p}(x), \\
    f^{-}_{p}(x) &= g^-_{p}(\vec{x})e^{ i\omega_p t}.
\end{align}
We will henceforth drop the $+$ superscript from the positive frequency modes. We have the Dirac current inner product, given by
\begin{equation} \label{eq: dirac inner product}
    \llangle f_{p}, f_{p'}\rrangle_{D} = i\int_{\Sigma} \overline{f}_{p}(x) \slashed{n} f_{p'}(x) \sqrt{h}\, d^{d-1}x.
\end{equation}
Here $n^\mu$ is the future directed normal to the chosen Cauchy surface of integration. We thus have $n_{\mu}=-\sqrt{\alpha(\vec{x})}\,\delta^{0}_{\mu}$. Via the conventional normalisation, we have $\llangle f_{p}, f_{p'}\rrangle_{D} = \delta^{d-1}(\vec{p}-\vec{p}\,')$.

We can see how to satisfy this normalisation condition straightforwardly. Inserting the basis elements, \eqref{Diracsolns}, into \eqref{eq: dirac inner product} we have 
\begin{equation}\label{eq: dirac mode orthogonality}
    \llangle f_{p}, f_{p'}\rrangle_{D} = e^{-i(\omega_{p'}-\omega_{p})t} \int_{\Sigma} g_{p}^{\dagger}(\vec{x})\, g_{p'}(\vec{x}) \sqrt{h}\, d^{d-1}x.
\end{equation}
We impose a condition on the $g_{p}$'s, such that they are normalised as
\begin{equation}
    \llangle g_{p},g_{p'}\rrangle_{\Sigma} := \int_{\Sigma} g_{p}^{\dagger}(\vec{x})\, g_{p'}(\vec{x}) \sqrt{h}\, d^{d-1}x = \delta^{d-1}(\vec{p}-\vec{p}\,'),
\end{equation}
and therefore our solutions form an orthonormal basis in the Dirac inner product. We also have that 
\begin{equation}
    \llangle g_{p},g^-_{p'}\rrangle_{D} = 0.
\end{equation}

 We can now consider the orthogonality of our basis, \eqref{Diracsolns}, with respect to the $L^2$ spacetime inner product, as we did for the bosons. This is 
\begin{equation}\label{eq: ferm ortho}
\begin{split}
    \llangle f_p, f_{p'} \rrangle_{L^2} =& \int \overline{f}_p(x)f_{p'}(x) \sqrt{-g}\, d^{d}x.\\
    =& \int e^{-i (\omega_{p'}-\omega_p)t}dt \int \sqrt{\alpha(\vec{x})}\, g^{\dagger}_p(\vec{x}) (-i\tilde{\gamma}^{0}) g_{p'}(\vec{x}) \sqrt{h}\, d^{d-1}x \\
    =& -2 \pi i \delta(\omega_p-\omega_{p'}) \int \sqrt{\alpha(\vec{x})}\, g^{\dagger}_p(\vec{x}) \tilde{\gamma}^{0} g_{p'}(\vec{x}) \sqrt{h}\, d^{d-1}x.
\end{split}
\end{equation}

Once again, we have taken the limit of the $t$ integral to $\pm \infty$. We see that the orthogonality condition we obtain is different from that of the Dirac inner product \eqref{eq: dirac mode orthogonality}. However, as before, positive and negative frequencies remain orthogonal in the $L^2$ inner product:
\begin{equation}\label{eq: ferm pos neg orthogonality}
    \llangle f_p, f^{-}_{p'} \rrangle_{L^2} = - 2 \pi i \delta(\omega_p+\omega_{p'}) \int \sqrt{\alpha(\vec{x})}\, g^{\dagger}_p(\vec{x}) \tilde{\gamma}^{0} g^-_{p'}(\vec{x}) \sqrt{h}\, d^{d-1}x=0,
\end{equation}
since $\omega_p, \omega_{p'} > 0$.

In the bosonic case, the analogous condition to \eqref{eq: ferm pos neg orthogonality} was sufficient to conclude that the SJ state is the same one that is picked out by the positive frequency modes, i.e. by $K^\mu$ and the Klein-Gordon inner product, due to the orthogonal support condition \eqref{eq: orthogonal support}. However, this is not sufficient in the fermionic case.\footnote{The orthogonal support condition, $WW^{T}=0$, leads to the fact that $\left(W+W^{T}\right)/2$ is unique. In the bosonic case, this is the state-dependent part of $W$, but in the fermionic case, this is the state-independent part of $W$. We therefore need an additional requirement to make the fermionic state-dependent part, $\left(W-W^{T}\right)/2$, unique.} The additional requirement, such that the SJ state is the same as the conventional state in static spacetimes, is that the matrix of inner products of solutions with the same frequency $\omega$ must be definite in the sign of the frequency. That is, for positive frequency solutions, the matrix $\left(\llangle f_p, f_{p'} \rrangle_{L^2}\right)_{p p'}$ must be positive definite, and for negative frequency solutions, the matrix $\left(\llangle f^{-}_p, f^{-}_{p'} \rrangle_{L^2}\right)_{p p'}$ must be negative definite, for all $p$, $p'$ such that $\omega_p=\omega_{p'}$. This requirement need not be met in general static spacetimes.

Note that this result, as well as the analogous one for the bosonic state, holds even in the (spacelike or null) boundary softened case. This is because the time integral in this case, given infinite limits, is unchanged.

\section{Entropy from the Algebra} \label{algebrasec}
In this appendix, we give an alternative, more algebraic, derivation of the entropy for a scalar field theory and a Majorana field theory in a region of spacetime. This derivation is of the type that was given by Sorkin in \cite{Sorkin2012} for a scalar theory, and we will begin by reviewing and generalising that formulation, taking a more constructive approach when possible.

 We observe that if a spacetime (region) is globally hyperbolic and therefore admits a Cauchy surface, then due to the quantum dynamics preserving the field algebra, the algebra associated to the spacetime is the same as that of its Cauchy surface.\footnote{This concept often goes by the name of primitive causality, or the time slice axiom.} We thus expect our algebraically formulated entropy, for a spacetime that is the domain of dependence of a Cauchy surface, to be  formally\footnote{to the extent that one can meaningfully identify the UV cutoffs (if required) in the spacetime and on the Cauchy surface.} equivalent to that associated with the Cauchy surface. Of course, we can also calculate entropies in spacetimes for which no Cauchy surface (or notion thereof) exists, but there the interpretation of this entropy is less clear.
 
\subsection{Bosonic Entropy} \label{appendix: algebraic S for bosons}

We proceed in the algebraic style, with local field operators $\phi(x)$.\footnote{These can be identified with $\phi(\vec{x})$ and $\dot{\phi}(\vec{x})$ on some Cauchy surface.} These form an algebra, $\mathfrak{U}$, under the relation $[\phi(x),\phi(x')]=i\Delta(x,x')$.

We define a state via its spacetime density matrix, $\rho$, which has the property $\text{Tr}(\rho A)=\langle A \rangle$ for all operators $A\in\mathfrak{U}$, where the angular brackets refer to an expectation value with respect to our state. The entropy is still defined via $S=-\text{Tr}\left(\rho \ln{\rho}\right)$.

To derive a formula for the entropy, we start with the spacetime Wightman function, and decompose it into its symmetric (anticommutator) and antisymmetric (commutator) parts, 
\begin{equation}
    W^{i j} = \langle \Psi|\phi^{i}\phi^{j}| \Psi \rangle = \frac{1}{2}\Bigl(\langle \Psi |\{\phi^{i},\phi^{j}\}| \Psi \rangle + \langle \Psi |[\phi^{i},\phi^{j}]| \Psi \rangle\Bigr) =: \frac{1}{2} (R^{ij} + i\Delta^{ij}),
\end{equation}
where the indices refer to spacetime elements. The Wightman function must be Hermitian, which ensures that both $R$ and $\Delta$ have real entries in the position basis.\footnote{Note that our $R$ differs by a factor of $2$ from the definition of the same symbol in \cite{Sorkin2012}.} As we are considering a bosonic theory, the symmetric part ($R$) is state-dependent while the antisymmetric part ($i\Delta$) is not.

Before we continue, we must impose that the Hilbert space is formed  of only solutions to the equation of motion. To do so, we use the fact that $\text{image}(\Delta)=\text{ker}(K)$ \cite{FGFTQF}, and  strip out the kernel of $i\Delta$ from the space. We also note that $W$ being positive semidefinite, and $R$ and $i\Delta$ being symmetric and antisymmetric respectively, implies that $\text{ker}(R) \subseteq \text{ker}(i\Delta)$.

With the kernel of $i\Delta$ removed, we now go to a $q$ and $p$ quadrature basis for the modes of $i\Delta$, wherein each mode has a single $q$ and $p$. This is nothing more than going to a Darboux basis \cite{Weinstein1973} for our nondegenerate (after the kernel removal) antisymmetric bilinear form $\Delta$, which always exists, but is not unique. This is a real basis, and hence must have twice the dimension of that of the $\phi$ basis. Having removed the kernel of $i\Delta$, we will have only one nontrivial commutation relation, $[q^{i},p^{j}] = i \delta^{i j}$. Choosing our basis to be $\{q^{1},\ldots,q^{n},p^{1},\ldots,p^{n}\}$, the Pauli-Jordan function takes the simple form
\begin{equation}
    \Delta = \begin{pmatrix}
        0 & \mathbb{I}_{n} \\
        - \mathbb{I}_{n} & 0
    \end{pmatrix}.
\end{equation}

Via Williamson's theorem \cite{WilliamsonThm, Adesso:2014npz}, we can go to a basis that diagonalises $R$, while keeping the form of $\Delta$, as it is a symplectic form. In this basis, $R$ is diagonal, with elements given by the positive eigenvalues of $i \Delta^{-1} R$, with a doubled multiplicity. In this basis, we see that our problem of calculating the system's entropy splits up into a series of independent calculations involving single mode oscillators, which have been Bogoliubov transformed into a diagonal form. These oscillators will each have a Wightman function 
\begin{equation}\label{eq: single mode W}
    W_{m} = \frac{1}{2}\left(R_{m}+i\Delta_{m}\right) = \frac{1}{2}\begin{pmatrix}
        \sigma_m & i \\
        -i & \sigma_m
    \end{pmatrix},
\end{equation}
where $\sigma_m\geq 1$. Here we are considering only a single decoupled mode, denoted by a subscript $m$. The eigenvalues of $-i\Delta^{-1}_{m} W_{m}$, which we label $\lambda_m^\pm$, are
\begin{equation}\begin{split}
    \text{spec}(-i\Delta^{-1}_{m}W_{m}) &= \frac{1}{2}\text{spec}(-i\Delta^{-1}_{m} R_{m} + \Delta^{-1}_{m}\Delta_{m}) \\
   & = \frac{1}{2}(\pm\sigma_m + 1) = \frac{1 \pm \sigma_m}{2}=:\lambda_m^\pm. \label{dubdub}
    \end{split}
\end{equation}
We can now calculate the entropy of our system. We start by considering a single mode. As our theory is Gaussian, the most general density matrix for it must be some exponentiated polynomial of at most quadratic order. Coupled with the fact that it must be Hermitian, we can write the density matrix as 
\begin{equation}
    \rho(q,q') = \sqrt{\frac{A}{\pi}}\exp\left(-\frac{A}{2}(q^{2}+q'^{2})-i\frac{B}{2}(q^{2}-q'^{2})-\frac{C}{2}(q-q')^{2})\right).
\end{equation}
Here we do not need to include linear terms, as they would result in $\langle\phi\rangle\neq0$.
We choose this specific form as it is amenable to finding the values of $A$, $B$, and $C$ via explicit calculation of the expectation values of quadratic products of $p$ and $q$, as we show below. The expectation values from $W_{m}$ can be read off from \eqref{eq: single mode W} and are
\begin{equation}
    \begin{split}
        \langle qq \rangle = \frac{\sigma_m}{2}, &\qquad \langle qp \rangle = \frac{i}{2},
        \\
        \langle pq \rangle = -\frac{i}{2}, &\qquad \langle pp \rangle = \frac{\sigma_m}{2}.
    \end{split}
\end{equation}
The expectation values from direct calculation are\footnote{Similar calculations were done in \cite{Chen:2020ild}.}:
\begin{equation}
\scalebox{0.97}{$\displaystyle
\begin{split}
    \frac{i}{2} &= \text{Tr}(qp\rho) \\
    &= \sqrt{\frac{A}{\pi}}\int dq\, dq' (-iq)\delta(q-q')\frac{\partial}{\partial q}\exp\left(-\frac{A}{2}(q^{2}+q'^{2})-i\frac{B}{2}(q^{2}-q'^{2})-\frac{C}{2}(q-q')^{2}\right) \\
    &= i(A-iB)\sqrt{\frac{A}{\pi}}\int dq (q^{2}) \exp\left(-Aq^{2}\right) = \frac{i}{2}\frac{A+iB}{A}.
\end{split}
$}
\end{equation}
From this, we see that $B=0$. We will not show the calculation of $\text{Tr}(pq\rho)$ explicitly, but one would get $-\frac{i}{2}+\frac{B}{2 A}$, retrieving the commutation relations when subtracting from $\text{Tr}(qp\rho)$. Setting $B=0$, we now calculate the rest of the expectation values:
\begin{equation}
    \frac{\sigma_m}{2} = \text{Tr}(qq\rho) = \sqrt{\frac{A}{\pi}}\int dq\, q^{2}\exp\left(-Aq^{2}\right) = \frac{1}{2A},
\end{equation}
\begin{equation}
\scalebox{0.99}{$\displaystyle
\begin{split}
    \frac{\sigma_m}{2} = \text{Tr}(pp\rho) &= \sqrt{\frac{A}{\pi}}\int dq\, dq' (-i)^{2}\delta(q-q')\frac{\partial^{2}}{\partial q^{2}}\exp\left(-\frac{A}{2}(q^{2}+q'^{2})-\frac{C}{2}(q-q')^{2}\right) \\
    &= \sqrt{\frac{A}{\pi}}\int dq \left(A + C - A^{2}q^{2}\right) \exp\left(-Aq^{2}\right) = \frac{A}{2} + C.
\end{split}
$}
\end{equation}
From the above two equations we see that
\begin{equation}
    A = \sigma_m^{-1}, \qquad C = \frac{1}{2}\left(\sigma_m-\sigma_m^{-1}\right).
\end{equation}
We hence have a density matrix for a single $(q,p)$ pair as 
\begin{equation}
    \rho(q,q') = \sqrt{\frac{\sigma_m^{-1}}{\pi}}\exp\left(-\frac{\sigma_m^{-1}}{2}(q^{2}+q'^{2})-\frac{\sigma_m-\sigma_m^{-1}}{4}(q-q')^{2})\right).\label{dens1}
\end{equation}

The form of the density matrix in (\ref{dens1}) is used in the early entanglement entropy calculation of \cite{BKLS86}. We will here show the calculation of the entropy, for the sake of completeness, and we will use a method in the style of \cite{Srednicki1993}, for its brevity.

We know that the von Neumann entropy, $S=-\text{Tr}\left(\rho \ln{\rho}\right)$, reduces to $\displaystyle S=-\sum_n P_{n} \ln{P_{n}}$, when $\rho$ is written in its basis of eigenfunctions, as
\begin{equation}
    \rho = \sum_{n} P_{n} \ket{f_{n}}\bra{f_{n}}.
\end{equation}
As we have $\rho$ in the $q$ basis, it is convenient to find the eigenfunctions and eigenvalues in this basis too. We must hence solve the equation 
\begin{equation}
\scalebox{0.89}{$\displaystyle
\begin{split}
    \rho\ket{f}=P\ket{f} &\implies \int dq' \rho(q,q') f(q') = P f(q) \\
    &\implies \int dq' \sqrt{\frac{\sigma_m^{-1}}{\pi}}\exp\left(-\frac{\sigma_m^{-1}}{2}(q^{2}+q'^{2})-\frac{\sigma_m-\sigma_m^{-1}}{4}(q-q')^{2})\right) f(q') = P f(q). \label{ef}
\end{split}
$}
\end{equation}

We can see that the density matrix is a Hilbert-Schmidt integral kernel, which is self-adjoint. This means that the eigenfunctions form a countable basis, and the eigenvalues satisfy
\begin{equation}
    \int dq\, dq' |\rho(q,q')|^{2} = \sum_{n}P_{n}^{2} \label{HS}
\end{equation}
(see e.g. \cite{Stone1932}). With integral operators, it is often easier to guess a solution and check that it is valid.  This is what was done for $\rho$ in \cite{Srednicki1993}, and it can be checked that the solution satisfies (\ref{ef}) and (\ref{HS}).  The eigenfunctions and eigenvalues are
\begin{equation}
\begin{aligned}
    f_{n}(q) &= H_{n}(q)\exp{\left( -\frac{q^{2}}{2}\right)}, \\
    P_{n} &= \left(1-\mu_m\right) \mu_m^{n}, \qquad \mu_m=\frac{\sigma_m-1}{\sigma_m+1}.
\end{aligned}
\end{equation}
Here $H_{n}(x)$ are the Hermite polynomials. Upon computation, we find that
\begin{equation}
\begin{aligned}
    \int dq\, dq' |\rho(q,q')|^{2} &= \frac{\sigma_m^{-1}}{\pi} \int dq\, dq' \exp\left(-\sigma_m^{-1}(q^{2}+q'^{2})-\frac{\sigma_m-\sigma_m^{-1}}{2}(q-q')^{2})\right) \\
    &= \frac{\sigma_m^{-1}}{\pi}\pi = \sigma_m^{-1},
\end{aligned}
\end{equation}
and
\begin{equation}
    \sum_{n}P_{n}^{2} = \sum_{n}\left(1-\mu_m\right)^{2} \mu_m^{2n} = \frac{(1-\mu_m)^{2}}{1-\mu_m^{2}} = \sigma_m^{-1},
\end{equation}
and hence (\ref{HS}) is satisfied. As this guarantees we have all the eigenvalues and eigenfunctions, we can then calculate the entropy via summing over the different $P_{n}$ in $\displaystyle - P_{n} \ln{P_{n}}$, which gives
\begin{equation}
    S_m = -\ln{(1-\mu_m)} - \frac{\mu_m}{1-\mu_m}\ln{\mu_m},
\end{equation}
which we then rewrite in terms of $\sigma_m$ to obtain
\begin{equation}\label{eq: S lambda}
    S_m =\frac{\sigma_m+1}{2}\ln{\frac{\sigma_m+1}{2}} - \frac{\sigma_m-1}{2}\ln{\frac{\sigma_m-1}{2}}.
\end{equation}
We note from the positive semidefiniteness of $W$ that $\sigma_m \geq 1$, and so the entropy of any mode is nonnegative. We now express \eqref{eq: S lambda} in terms of the eigenvalues of the decoupled $-i\Delta^{-1}_{m}W_{m}$, as in  (\ref{dubdub}), which gives 
\begin{equation}
    S_m =\lambda_m^+\ln{(\lambda_m^+)} + \lambda_m^-\ln{(-\lambda_m^-)}.
\end{equation}
Having covered the single mode case, the full entropy is simply the sum over the entropies of each mode. Considering the full operator $-i\Delta^{-1} W$, with eigenvalues $\lambda$, and noting that $\sigma_m \geq 1$, we have the total entropy
\begin{equation} \label{eq: S lambda algebra}
    S = \sum_\lambda \lambda \ln{|\lambda|},
\end{equation}
where, as mentioned before, the kernel of $i\Delta$ has been removed. This can be further emphasised by obtaining the eigenvalues as the solutions to a generalised eigenvalue equation,
\begin{equation}\label{boson_ev}
    W v = i \lambda \Delta v,
\end{equation}
where it is understood that we cannot have $\Delta v =0$, i.e. that the kernel of $\Delta$ is necessarily removed. Equations \eqref{eq: S lambda algebra} and \eqref{boson_ev} are the same as \eqref{Ssum} and \eqref{Sev}, that we derived in Section \ref{bosonsec}. We have thus arrived at the same result in an algebraic manner, as we set out to do. 
\subsection{Fermionic Entropy}\label{appendix: algebraic S for fermions}
We now repeat the above treatment for a Majorana field. We begin with the fermionic version of the Wightman function, 

\begin{equation}
    W^{i j} = \langle \Psi|\psi^{i}\bar\psi^{j}| \Psi \rangle = \frac{1}{2}\Bigl(\langle \Psi |\{\psi^i, \bar{\psi}^j\}| \Psi \rangle + \langle \Psi |[\psi^{i},\bar{\psi}^{j}]| \Psi \rangle\Bigr) =: \frac{1}{2} (  i\Delta_F^{ij}+R_F^{ij}).
\end{equation}
As we are considering a fermionic theory, the ``antisymmetric" part ($R_F$) is state-dependent while the ``symmetric" part ($i\Delta_F$) is not.

We then write $W$ in a linearly independent basis of modes, remove the kernels of $i\Delta_F$ and $R_F$, and re-express it in the ``quadrature" basis for our modes. Here the equivalent to our previous $q$'s and $p$'s are elements of the Majorana basis $\gamma^i$ (not to be confused with the Dirac gamma matrices) \cite{ElliottFranz}. We have
\begin{equation}
    \{\gamma^{i},\gamma^{j}\}=\delta^{ij},
\end{equation}
and hence
\begin{equation}
   i \Delta_F = \begin{pmatrix}
        \mathbb{I}_{n} & 0 \\
        0 & \mathbb{I}_{n}
    \end{pmatrix} = \mathbb{I}_{2n}.
\end{equation}
We would like to write $W$ in a similar form to \eqref{eq: single mode W}, which requires preserving the diagonal form of $i\Delta_F$, while transforming $R_F$ into a skew block diagonal form. We can use some $O \in SO(2n)$ to transform $R_F$ into this form. We then have
\begin{equation}
    O^{T}R_F O = i\begin{pmatrix}
        0 & \Sigma \\
        -\Sigma & 0
    \end{pmatrix},
\end{equation}
where $\Sigma$ is diagonal, with entries $\sigma \in \mathbb{R}$. These entries  have the property that $\pm i \sigma$ are the eigenvalues of $R_F$. Once again, our Wightman function can be decomposed into contributions from each mode $m$ (with its basis $\gamma^1_m$, $\gamma^2_m$), for which we can write 
\begin{equation}
    W_{m} = \frac{1}{2}\left(i\Delta_{F,{m}}+R_{F,m}\right) = \frac{1}{2}\begin{pmatrix}
        1 & i\sigma_m \\
        -i\sigma_m & 1
    \end{pmatrix}.
\end{equation}
Keeping with our treatment of the bosonic case, we will next find the density matrix in a representation. For a single mode, the density matrix is very simple, due to the Clifford nature of the $\gamma$ operators causing the exponential to truncate. However, instead of parameterising with exponentiated constants, it is simpler still to write $\rho$ as the most general Hermitian matrix we can form.\footnote{One can also go through the same process as with the bosons, if one prefers. We would then have the density matrix in exponential form, which simplifies to $\rho = \frac{1}{2}\left(\mathbb{I}-i \text{tanh}(\kappa) \gamma^{1}_m\gamma^{2}_m\right)$, since the only quadratic term is of the form $\gamma^{1}_m\gamma^{2}_m$. We would find $\kappa = \text{tanh}^{-1}(\sigma_{m})$, and thus get the same result for the entropy.}

We now choose an explicit representation for our Majorana operators, namely 
\begin{equation}
    \gamma^{1}_m = \frac{1}{\sqrt{2}}\begin{pmatrix}
        0 & 1 \\
        1 & 0 \\
    \end{pmatrix},
    \quad
    \gamma^{2}_m = \frac{1}{\sqrt{2}i}\begin{pmatrix}
        0 & 1 \\
        -1 & 0 \\
    \end{pmatrix}.
\end{equation}
Our density matrix will thus be a $2\times 2$ Hermitian matrix of the form 
\begin{equation}
    \rho = \begin{pmatrix}
        A & B \\
        B^* & C
    \end{pmatrix}.
\end{equation}

Via $\text{Tr}(\gamma^{i} \rho)=0$, we see that $B=0$. We can then calculate the expectation values in the same way as in the bosonic case, and find
\begin{equation}
    \frac{1}{2} = \text{Tr}(\gamma^{1}_m\gamma^{1}_m\rho) = \frac{1}{2}(A+C),
\end{equation}
\begin{equation}
    \frac{i \sigma_m}{2} = \text{Tr}(\gamma^{1}_m\gamma^{2}_m\rho) = \frac{i}{2}(A - C).
\end{equation}
The other two equations tell us nothing new. From this, we see that
\begin{equation}
    A = \frac{1+\sigma_m}{2}, \quad C = \frac{1-\sigma_m}{2}, \qquad \text{and therefore}\qquad \rho = 
    \frac{1}{2} \begin{pmatrix}
        1+\sigma_m & 0 \\
        0 & 1-\sigma_m
    \end{pmatrix}.
\end{equation}
The von Neumann entropy for a single mode can now be read off as
\begin{equation}
S_m = -\left(\frac{1+\sigma_m}{2}\ln{\frac{1+\sigma_m}{2}} + \frac{1-\sigma_m}{2}\ln{\frac{1-\sigma_m}{2}}\right).
\end{equation}

Let us now, in analogy with the bosonic case, consider the eigenvalues of $-i\Delta_{F,m}^{-1}W_{m}$, $\lambda_m^{+}$ and $\lambda_m^{-}$. In this basis $i\Delta_{F,m} = \mathbb{I}_2$, and so its presence may seem trivial in the expression $-i\Delta_{F,m}^{-1} W_m$. It is, however, needed for bookkeeping purposes, as we rescaled the quadrature expectation values, and so there would be additional scaling factors contributed by the presence of $-i\Delta_{F,m}^{-1}$ when changing basis. The eigenvalues are $\left\{\frac{1}{2}(1+\sigma_m), \frac{1}{2}(1-\sigma_m)\right\}$, and therefore
\begin{equation}
    S_m=-\lambda_m^{+}\ln{\lambda_m^{+}} - \lambda_m^{-}\ln{\lambda_m^{-}},
\end{equation}
where the positive semidefiniteness of $W$ guarantees that $0\leq\lambda_m \leq 1$. As before, the full entropy is the sum over all the single mode entropies. Considering the full operator $-i\Delta_F^{-1}W$, with eigenvalues $\lambda$, we can write the total entropy as
\begin{equation} 
    S = -\sum_\lambda \lambda \ln\lambda \label{ferm_S}.
\end{equation} 

We can once again recast this problem to get the eigenvalues by solving the generalised eigenvalue equation
\begin{equation}\label{fermion_ev}
    W v = i\lambda \Delta_F\, v.
\end{equation}
When written in this form, it is understood that the kernel of $\Delta_F$ is always to be removed, i.e. solutions with $\Delta_F\, v=0$ are disallowed.  Equations \eqref{ferm_S} and \eqref{fermion_ev} are the same as \eqref{Ssumf} and \eqref{eq: gen eig ferm}, respectively, that we derived in Section \ref{fermsec}. Once again, we have obtained our previous result in a more algebraic manner, as intended. 

\section{Modifications to $K$ and $\Delta$ in a Discrete Setting}\label{ap: volume law}
In this appendix, we speculate about the source of the  eigenfunctions with small eigenvalues observed and truncated in Section \ref{sec: ee example}, based on their qualitative features.

\begin{figure}[ht]
  \centering
  \begin{tikzpicture}[scale=0.85, transform shape]
    \node[anchor=south west, inner sep=0] (mainfig)
      at (0,0)
      {\includegraphics[width=1.0\linewidth]{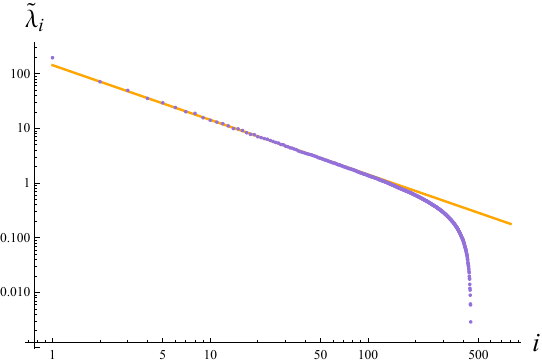}};
    \begin{scope}[x={(mainfig.south east)}, y={(mainfig.north west)}]
      \node[anchor=south west]
        (inset1)
        at (0.1,0.15)
        {\includegraphics[width=0.4\linewidth]{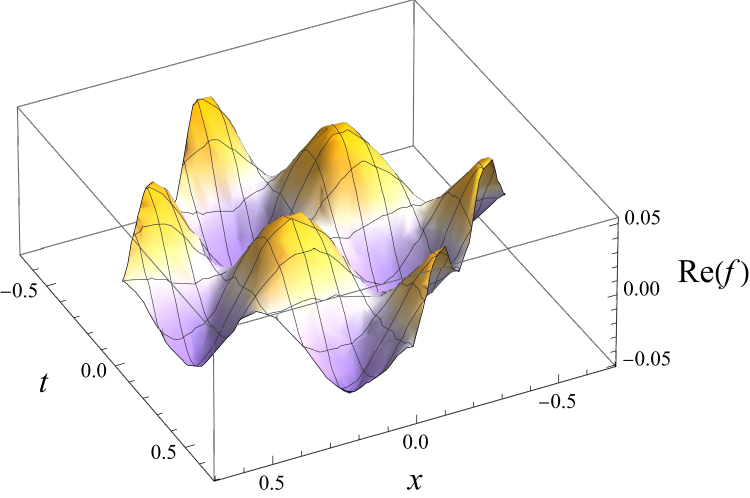}};
      \node at (inset1.center) {
        \phantomsubcaption\label{fig: big eigenfunction}
      };

      \node[
        anchor=north west,
        font=\bfseries\small,
        xshift=15pt,
        yshift=20pt
      ]
        at ([yshift=-5pt, xshift=-55pt]inset1.south east)
        {(a)};
      \node[anchor=north east]
        (inset2)
        at (1.07,0.95)
        {\includegraphics[width=0.4\linewidth]{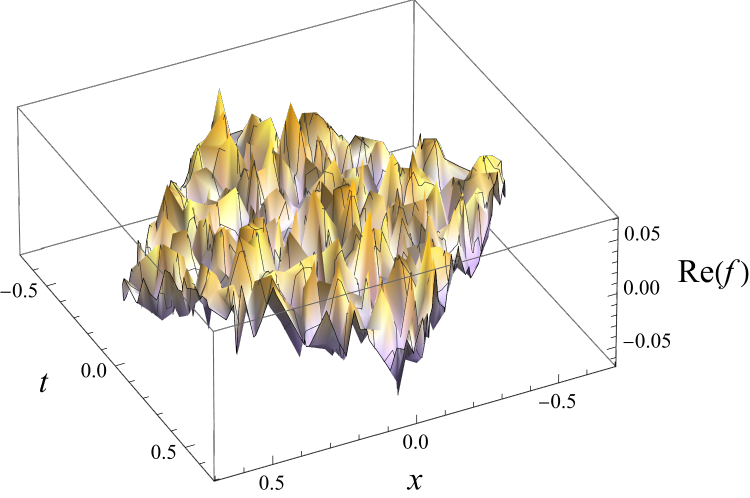}};

      \node at (inset2.center) {
        \phantomsubcaption\label{fig: middle eigenfunction}
      };
        
      \node[
        anchor=north west,
        font=\bfseries\small,
        xshift=-40pt,
        yshift=15pt
      ]
        at (inset2.south east)
        {(b)};
      \draw[
        color=black, 
        line width=0.75pt
      ]
        (0.25,0.54)
        -- (0.2715,0.72);
      \draw[
        color=black,
        line width=0.75pt
      ]
        (0.796, 0.536)
        -- (0.787,0.454);

    \end{scope}
  \end{tikzpicture}

  \caption{A log-log plot of the positive eigenspectrum of $i\Delta$ in a $1+1$d causal diamond. The causal set eigenvalues from a $900$-element sprinkling, into a diamond of side length $1$, are shown in purple. The asymptotic power law expected from the continuum eigenvalues \eqref{eq: spectral density}, rescaled by a factor of $\frac{1}{2}\tilde\rho=450$, is shown in orange (the factor of half is necessary to account for the two families of eigenfunctions). (a) A first order (Delaunay) interpolation of the real part of the eigenfunction corresponding to the $4^{\text{th}}$ largest positive eigenvalue of $i\Delta$. This eigenfunction shows the typical characteristics of a continuum eigenfunction. (b) A first order (Delaunay) interpolation of the real part of the eigenfunction corresponding to the $225^{\text{th}}$ largest positive eigenvalue. This eigenfunction shows the typical characteristics of a non-continuumlike eigenfunction.}  \label{fig: loglogeigenvalues}
\end{figure}
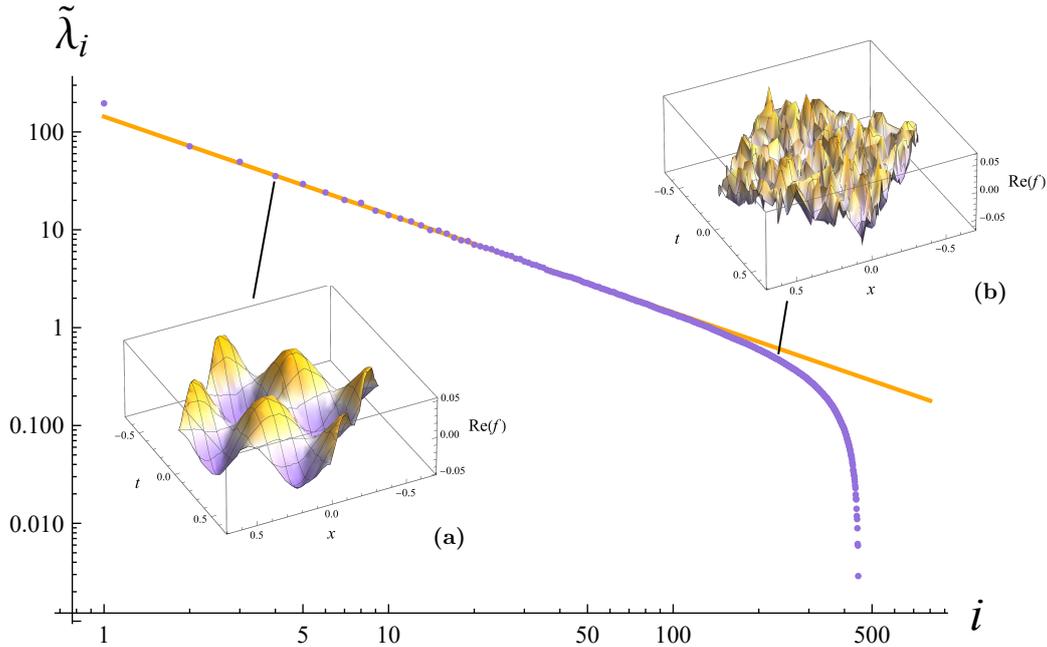

As mentioned in Section \ref{sec: ee example}, the spectrum of $i\Delta$ in the causal set has a large  (order $N$) number of small values, with corresponding eigenfunctions that have no continuum counterpart, that we would like to exclude. These are shown in the tail of Figure \ref{fig: loglogeigenvalues}, along with an example corresponding eigenfunction in Figure \ref{fig: middle eigenfunction}. For comparison, an example continuumlike eigenfunction corresponding to a large eigenvalue is shown in Figure \ref{fig: big eigenfunction}.

Note that all eigenfunctions corresponding to these small eigenvalues resemble the one in Figure \ref{fig: middle eigenfunction}. Namely, they are nearly all noise-like, with features only at the discreteness scale. Therefore, while the eigenvalues can be ordered based on their magnitude, there is no noticeable scale hierarchy based on any feature of the eigenfunctions. It is thus not surprising, if each such eigenfunction contributes a roughly equal and large entropy, and owing to the order $N$ number of them, this gives rise to a spacetime volume scaling law. 
We previously explained the truncation procedure which eliminates these degrees of freedom from our entropy calculation. Here, we try to understand their presence in terms of a modified equation of motion.

In the discussion that follows, $K$ and $\Delta$ will abstractly correspond to their causal set versions that describe a local field theory, analogous to those of the continuum, but regulated by discreteness. $\tilde K$ and $\tilde \Delta$ will stand for more general modifications, as we describe below. The causal set $\Delta$ we start with in Section \ref{sec: ee example} would correspond to $\tilde\Delta$ in this appendix.

Let us consider the effect of slightly modifying the causal propagator $\Delta$, such that that which was in its kernel now maps to a small nonzero value. We would then have
\begin{equation}
    \tilde \Delta = \Delta + \epsilon A, \quad \epsilon \in \mathbb{R}, \quad A^\dagger = -A.
\end{equation}
Note that we modified $\Delta$ in such a way that it remains antisymmetric, so that $i\tilde\Delta$ is still a sensible commutator. In our case, $A$ could be a random antisymmetric matrix, related to the randomness of the Poisson sprinkling.

Now consider the modified relation between $K$ and $\tilde \Delta$,
\begin{equation}
    K \tilde \Delta= K(\Delta + \epsilon A) = \epsilon K A =: B.
\end{equation}
The original causal propagator had $B = 0$, but now we have a $B$ that is generically nonzero. This means that we no longer have $K \varphi = 0$, i.e. our original equation of motion does not hold. We can express the modification to the equation of motion as the addition of a source field term, $J$, where
\begin{equation}\label{eq: sourced eom}
    J := K \varphi = K \tilde{\Delta} f = \epsilon K A f\neq 0.
\end{equation}
The support of $J$ is the same as the support of $A$ (which in the causal set is everywhere), as $K$, being local (in the continuum it is a differential operator), does not enlarge support.

We can rewrite \eqref{eq: sourced eom} as a new homogeneous equation of motion,
\begin{equation}
    \tilde K \varphi = 0,
\end{equation}
where $\tilde K := \mathcal Q K$, and $\mathcal QJ=0$. In general, we would expect the operator $\mathcal Q$, and therefore $\tilde K$, to be  nonlocal in space and time. This expectation is because the $J$'s in the causal set lack structure, and therefore the $\mathcal Q$ annihilating them ($\mathcal QJ=0$) is likely to be an integral or sum operator involving relations at several or even all elements. 
It has in fact already been observed that nonlocal scalar field theories naturally arise in causal set theory \cite{Sorkin:2007qi, Benincasa:2010ac, Aslanbeigi:2014zva}.

Viewed in this way, the spectrum truncation of Section \ref{sec: spectrum truncation} may also be interpreted as removing degrees of freedom that have an entirely nonlocal nature. Similar, though subleading, modifications to the remaining degrees of freedom are also expected, and these may be responsible for the extra entropy obtained in Section \ref{sec: ee results}.

\bibliographystyle{JHEP}

\bibliography{biblio.bib}
\end{document}